\documentclass[twocolumn]{aastex7}

\usepackage{booktabs}

\usepackage{multirow}
\usepackage{comment}
\usepackage{makecell}
\usepackage{amsmath,amssymb}

\usepackage{caption}
\usepackage{subcaption}

\shorttitle{X-ray and multiwavelength study of PWN G0.9+0.1}
\shortauthors{G. Brunelli et al}

\received{Month DD, YYYY}
\revised{Month DD, YYYY}
\accepted{\today}

\submitjournal{ApJ}

\begin{document}

\title{New Hard X-Ray and Multiwavelength Study of the PeVatron Candidate PWN G0.9+0.1 in the Galactic Center Region}

\author[orcid=0009-0008-2078-2456]{Giulia Brunelli}
\affiliation{Dipartimento di Fisica e Astronomia (DIFA) Augusto Righi, Università di Bologna, via Gobetti 93/2, I-40129 Bologna, Italy}
\affiliation{INAF - Osservatorio di Astrofisica e Scienza dello spazio di Bologna, Via Piero Gobetti 93/3, 40129 Bologna, Italy}
\email[show]{giulia.brunelli@inaf.it}  

\author[orcid=0000-0002-9709-5389]{Kaya Mori}
\affiliation{Columbia Astrophysics Laboratory, 550 West 120th Street, New York, NY 10027, USA}
\email{kaya@astro.columbia.edu}  

\author[orcid=0000-0002-9103-506X]{Jaegeun Park}
\affiliation{Department of Astronomy and Space Science, Chungbuk National University, Cheongju, 28644, Republic of Korea}
\email{geunjaep@gmail.com} 

\author[orcid=0000-0001-9633-3165]{Jordan Eagle}
\affiliation{NASA Goddard Space Flight Center, Greenbelt, MD 20771, USA}
\email{jordanlynneagle@gmail.com} 

\author[orcid=0000-0002-4441-7081]{Moaz Abdelmaguid}
\affiliation{New York University Abu Dhabi, PO Box 129188, Abu Dhabi, UAE}
\email{m.abdelmaguid@nyu.edu} 

\author[orcid=0000-0002-3310-1946]{Melania Nynka}
\affiliation{Kavli Institute For Astrophysics and Space Research, Massachusetts Institute of Technology, Cambridge, MA, USA}
\email{mnynka@mit.edu} 

\author[orcid=0000-0002-6389-9012]{Hongjun An}
\affiliation{Department of Astronomy and Space Science, Chungbuk National University, Cheongju, 28644, Republic of Korea}
\email{hjan@cbnu.ac.kr} 

\author[orcid=0000-0003-0890-4920]{Aya Bamba}
\affiliation{Department of Physics, University of Tokyo, Tokyo 113-0033, Japan}
\email{bamba@phys.s.u-tokyo.ac.jp} 

\author[orcid=0000-0003-4679-1058]{Joseph D. Gelfand}
\affiliation{New York University Abu Dhabi, PO Box 129188, Abu Dhabi, UAE}
\email{jg168@nyu.edu}

\author[orcid=0000-0003-0293-3608]{Gabriele Ponti}
\affiliation{INAF – Osservatorio Astronomico di Brera, Via E. Bianchi 46, 23807 Merate, Italy}
\affiliation{Max-Planck-Institut für extraterrestrische Physik, Gießenbachstraße 1, 85748 Garching, Germany}
\affiliation{Como Lake Center for Astrophysics (CLAP), DiSAT, Università degli Studi dell'Insubria, via Valleggio 11, 22100 Como, Italy}
\email{gabriele.ponti@inaf.it}

\author[orcid=0000-0001-6189-7665]{Samar Safi-Harb}
\affiliation{Department of Physics and Astronomy, University of Manitoba, Winnipeg, MB R3T 2N2, Canada}
\email{samar.safi-harb@umanitoba.ca}

\author[orcid=0000-0001-8202-9381]{Vito Sguera}
\affiliation{INAF - Osservatorio di Astrofisica e Scienza dello spazio di Bologna, Via Piero Gobetti 93/3, 40129 Bologna, Italy}
\email{vito.sguera@inaf.it}

\author[orcid=0000-0002-8853-9611]{Cristian Vignali}
\affiliation{Dipartimento di Fisica e Astronomia (DIFA) Augusto Righi, Università di Bologna, via Gobetti 93/2, I-40129 Bologna, Italy}
\affiliation{INAF - Osservatorio di Astrofisica e Scienza dello spazio di Bologna, Via Piero Gobetti 93/3, 40129 Bologna, Italy}
\email{cristian.vignali@unibo.it}

\author[orcid=0009-0001-6471-1405]{Jooyun Woo}
\affiliation{Columbia Astrophysics Laboratory, 550 West 120th Street, New York, NY 10027, USA}
\email{jw3855@columbia.edu}

\author[orcid=0000-0001-6320-1801]{Roberta Zanin}
\affiliation{Cherenkov Telescope Array Observatory gGmbH, via Piero Gobetti 93/3, 40129 Bologna, Italy}
\email{roberta.zanin@cta-observatory.org}

\begin{abstract}
We present a new X-ray study and multiwavelength spectral energy distribution (SED) modeling of the young pulsar wind nebula (PWN) powered by the energetic pulsar PSR~J1747--2809, inside the composite supernova remnant (SNR) G0.9+0.1, located in the Galactic Center region. The source is detected by NuSTAR up to 30\,keV with evidence for the synchrotron burnoff effect in the changing spatial morphology with increasing energy. The broadband 2--30\,keV spectrum of PWN~G0.9+0.1 is modeled by a single power law with photon index $\Gamma=2.11\pm0.07$. We combined the new X-ray data with the multiwavelength observations in radio, GeV, and TeV gamma rays and modeled the SED, applying a one-zone and a multi-zone leptonic model. The comparison of the models is successful, as we obtained physically compatible results in the two cases. Through the one-zone model, we constrain the age of the system to $\sim2.2$\,kyr, as well as reproduce the observed PWN and SNR radio sizes. In both the one-zone and multi-zone leptonic models, the electron injection spectrum is well-described by a single power law with spectral index $p \sim 2.6$ and a maximum electron energy of $\sim2$\,PeV, suggesting the source could be a leptonic PeVatron candidate. We estimate the average magnetic field to be $B_{\rm PWN} \sim 20\ \mu$G. We also report the serendipitous NuSTAR detection of renewed X-ray activity from the very faint X-ray transient XMMU~J174716.1--281048 and characterize its spectrum.
\end{abstract}


\section{Introduction} \label{sec:introduction}
Pulsar wind nebulae (PWNe) are bubbles of relativistic particle outflows, primarily composed of electrons and positrons that are continuously injected from energetic pulsars. They usually show multiwavelength nonthermal spectra, shining from radio to X-rays through synchrotron radiation from interactions between the relativistic particles and the PWN magnetic field, and in the gamma rays through inverse Compton scattering (ICS) of the relativistic particles off the surrounding photon fields, such as the cosmic microwave background (CMB) and local infrared and optical interstellar radiation. ICS, in particular of the CMB and infrared photons, is responsible for the very-high-energy (VHE, $E\geq50$~GeV) emission of PWNe, which has been detected by imaging atmospheric Cherenkov telescope (IACT) arrays \citep[e.g.,][]{HESS_2018}. Recent observations from the Large High Altitude Air Shower Observatory (LHAASO) and the High-Altitude Water Cherenkov Observatory (HAWC) revealed an increasing number ($\sim15$ reported by LHAASO, $\sim10$ by HAWC) of ultra-high-energy (UHE, $E>100$\,TeV) gamma-ray sources associated with PWNe \citep{Albert_2020, Cao_2024}, making them prime candidates for Galactic ``PeVatrons" responsible for the acceleration of leptons with energies above 1\,PeV in the Galaxy.

Since the emission of PWNe can span multiple decades in energy, it is crucial to conduct multiwavelength observations to perform an accurate modeling of their spectral energy distribution (SED). This also allows us to determine the spatial and energy distributions of relativistic leptons accelerated at the termination shock (TS) and how the supernova remnant (SNR) interior and surrounding interstellar medium (ISM) influence the PWN. The SED of PWNe can dramatically change with time due to interactions with the SNR and the structure of the ISM. There are three main stages of the evolution of a PWN: the free expansion phase ($t_{\rm age}\lesssim 10$\,kyr), the reverberation phase ($10 \lesssim t_{\rm age}\lesssim 100$\,kyr), and the post-reverberation phase ($t_{\rm age}\gtrsim 100$\,kyr). For complete reviews on the different evolutionary stages, refer to \cite{Mitchell_2022, Olmi_2023}.
Different models, such as \cite{Gelfand_2009}, \cite{Tanaka_Takahara_2011}, or \cite{Torres_2014}, have been developed to study how the SED of a PWN evolves with time and simultaneously constrain the properties of the system, such as the age or the electron spectrum of the pulsar wind, and reproduce the present-day SED. A different approach for the SED modeling is that of multi-zone models, which assume that some of the PWN properties are changing with the radius, but do not consider evolutionary changes. An example is the model developed by \cite{Kim_2020}, which has the goal of reproducing the present-day SED of PWNe as well as their observed spatially-dependent X-ray features (photon index and surface brightness) with varying distance from the pulsar.

One of the key parameters to constrain from SED modeling is the maximum energy ($E_{\rm max}$) particles are accelerated. A source is established as a PeVatron candidate if $E_{\rm max} \gtrsim 1\,$PeV. VHE observations alone cannot put stringent constraints on $E_{\rm max}$ due to the Klein-Nishina cutoff that affects the ICS emission at TeV--PeV energies. However, the highest-energy particles (GeV--PeV energies) are responsible for the hard X-ray (from tens of keV up to MeV) emission of PWNe, which does not suffer from the Klein-Nishina effect. The Nuclear Spectroscopic Telescope Array (NuSTAR) is sensitive to the wide energy range from 3 to 79~keV and is one of the best instruments to study PWNe in the hard X-rays. Combined with soft X-ray instruments such as Chandra or XMM--Newton, working at energies lower than 10\,keV, it is able to provide the best constraints on $E_{\rm max}$. An extensive NuSTAR observation campaign of seven PWNe associated with TeV sources resulted in the detection of all of them above 10\,keV \citep[e.g.,][]{Burgess_2022, Woo_2023, Park_2023a, Park_2023b, Kim_2024, Pope_2024}. Subsequent multiwavelength SED studies find that many of them are PeVatron candidates with $E_{\rm max}$ exceeding 1\,PeV, e.g., the ``Dragonfly" PWN \citep{Woo_2023} and the ``Eel" PWN \citep{Burgess_2022}. Nevertheless, the sample of sources that have been studied in this context is still limited. A larger number of PWNe, with different ages and energetics, needs to be investigated to better understand what conditions allow the sources to accelerate particles up to PeV energies. The NuSTAR observational campaign covered a variety of middle-aged PWNe, i.e., systems where the estimated characteristic age of the pulsar is of the order of tens of kiloyears. The exception is PWN~G0.9+0.1, which pulsar has a characteristic age of $\tau_c \sim 5$\,kyr.

The parent SNR of PWN~G0.9+0.1, located in the vicinity of the Galactic Center (GC), was reported for the first time by \cite{Kesteven_1968}. Later, \cite{Helfand_1987} classified it as a composite system having an SNR shell and a compact PWN at its center. The most recent study of the PWN in the radio band is presented in \cite{Dubner_2008}, where they analyzed Australia Telescope Compact Array (ATCA) and Very Large Array (VLA) observations at different frequencies to characterize the morphology and spectrum of the radio emission. They found a flat spectral index of $\alpha=-0.18 \pm 0.04$ by combining data at wavelengths of 3.6, 6, 20, and 90\,cm. No evidence of significant spectral variations in different regions of the PWN is found. They also provided a lower limit to the age of the system of $t\sim 1.1$\,kyr, obtained by comparing the total energy of the system and the estimated spindown power of the pulsar. The source was also detected by MeerKAT in the 1.28\,GHz mosaic of the GC \citep{MeerKAT_2022}. The PWN was first detected in the X-rays by BeppoSAX \citep{Mereghetti_1998, Sidoli_2000} as a hard and highly absorbed source, compatible with a location close to the GC. The study of \cite{Sidoli_2000} estimated the age of the system to be $t\sim2.7$\,kyr, using the same technique as \cite{Dubner_2008}, applied in this case to X-ray data. Follow-up observations with Chandra \citep{Gaensler_2000} revealed a torus-jet morphology resembling that of other young PWNe such as the Crab Nebula or Vela \citep{Bamba_2010}. Observations with XMM--Newton \citep{Porquet_2003, Holler_2012} found a spectral softening with increasing distance from the central pulsar, with the photon index varying from $\Gamma \simeq 1.2-1.4$ in the inner regions to $\Gamma \simeq 2.2-2.4$ in the outer regions. PWN~G0.9+0.1 is also detected in the gamma-rays as the second brightest VHE source in the GC, discovered for the first time by the High Energy Stereoscopic System (HESS) above 200\,GeV \citep{HESS_2005}. The PWN has been detected later by both the Very Energetic Radiation Imaging Telescope Array System \citep[VERITAS,][]{VERITAS_2016, VERITAS_2021} and the Major Atmospheric Gamma Imaging Cherenkov (MAGIC) telescopes \citep{MAGIC_2017, MAGIC_2020}. At teraelectronvolt energies, the source has a power-law spectrum with a photon index of $\Gamma \sim 2$ and shows no apparent signs of a cutoff at energies up to 20\,TeV. Furthermore, HAWC detected a UHE source near the position of G0.9+0.1 \citep{HAWC_2024}, but offset from the X-ray and TeV position of PWN~G0.9+0.1 by $\sim0.5^\circ$ and with a significance below $5\sigma$. 

The pulsar powering PWN~G0.9+0.1, PSR~J1747--2809, was first detected by \cite{Camilo_2009} at 2\,GHz using the NRAO Green Bank Telescope. They estimated the spin period and its derivative to be $P=52$ ms and $\dot{P}\simeq1.6 \times 10^{-13}$\,s\,s$^{-1}$, from which they derived a characteristic age of $\tau_c\simeq 5.3$\,kyr and a spindown power of $\dot{E}\simeq 4.3 \times 10^{37}$ erg s$^{-1}$, making it one of the most energetic Galactic pulsars ever detected. \citet{Camilo_2009} obtained a dispersion measure of $DM = 1133$\,pc cm$^{-3}$ and an associated distance of $d \sim 13$\,kpc. The estimate for the distance, derived using the electron density model NE2001 from \cite{Cordes_Lazio_2002}, is highly uncertain ($0.8 \le d_{10} \le 1.6$, where $d_{10}$ is the distance in units of 10\,kpc) and the authors could not exclude the possibility of the pulsar being located close to the GC ($d \sim 8-9$\,kpc). The pulsar age was estimated to be $\sim 2-3$\,kyr. No pulsations have been detected so far in the X-rays or gamma-rays, though Chandra observations identified a hard X-ray point source named CXOU~J174722.8--280915 that is likely to be the pulsar. \cite{Gaensler_2000} estimated the X-ray luminosity of the pulsar source to be $\leq 1$ \% of the PWN luminosity. PSR~J1747--2809 deviates from the typical $L_{\rm X, PWN}\sim (1-10) \times L_{\rm X, PSR}$ law observed for X-ray PWNe \citep{Li_2008}, but it is not an isolated case. Other PWNe lack bright X-ray or pulsed emission from their associated pulsars, such as the PWNe inside SNR~G16.7+0.1 \citep{Chang_2018} of $\sim2$\,kyr age, or SNR~G63.7+0.1 \citep{Matheson_2016}, with an age of $\sim8$\,kyr, or the older SNR~G327.1--1.1 \citep{Temim_2015}.

Multiwavelength SED studies have been performed, such as the one-zone model reported in \cite{HESS_2005}, through which they derived an equipartition magnetic field of $\approx5\ \mu$G for a distance of $d=8.5$\,kpc, a source radius of $r=1'$ and fixing $E_{\rm max}=500$~TeV. A different approach was used by \cite{Fang_Zhang_2010}, who adopted an injection spectrum composed of a low-energy Maxwellian distribution and a high-energy power law tail, and derived a maximum energy of $E_{\rm max}\sim900$~TeV and a magnetic field of the PWN of $8.1\ \mu$G. Both \cite{Tanaka_Takahara_2011} and \cite{Torres_2014} tested a one-zone time-dependent model of a spherical PWN, with slightly different assumptions, and investigated two scenarios for the distance to the source, $d_1=8.5$\,kpc and $d_2=13$\,kpc, deriving maximum energies in the range $E_{\rm max}\sim 400 - 900$\,TeV and magnetic fields $B_{\rm PWN} \sim 10-20\ \mu$G, depending on the model. \cite{Fiori_2020} applied a similar model with simulated Cherenkov Telescope Array Observatory (CTAO) data and derived maximum energies exceeding 600\,TeV, but assuming a distance of 13\,kpc.

Despite the several attempts, $E_{\rm max}$ is not well constrained in the absence of hard X-ray or MeV gamma-ray data. Previous works have overall estimated $E_{\rm max}$ in the range 400--900\,TeV. Our work, including both NuSTAR data and Fermi--LAT upper limits, aims to improve the previous SED studies and provide better constraints on the maximum energy that particles produced by PSR~J1747--2809 can reach. The characteristics of PWN~G0.9+0.1 in both X-rays and VHE strongly suggest it is likely a PeVatron, and the goal of our work is to investigate this hypothesis. We present our broadband X-ray analysis of the PWN, including both archival XMM--Newton data and new NuSTAR observations, in Sect.~\ref{sec:xr_obs_analysis}. We employ the dynamical one-zone model from \cite{Gelfand_2009} and the multi-zone model by \cite{Kim_2020}, including multiwavelength data from radio to TeV bands, presented in Sect.~\ref{sec:sed}. The comparison between the two model results and our interpretation are discussed in Sect.~\ref{sec:discussion}.

\section{X-Ray Observations and Data Analysis} \label{sec:xr_obs_analysis}
\subsection{X-Ray Observations} \label{subsec:xr_obs}
We analyzed two archival XMM--Newton observations targeting PWN~G0.9+0.1, the first (Obs ID: 0112970201, 18\,ks exposure time) taken in 2000 and the second (Obs ID: 0144220101, 52\,ks exposure time) in 2003 (see Tab.~\ref{tab:xray_obs}). The source is in the field of view (FoV) of several additional XMM--Newton observations; however, we analyzed only the data sets where the source was observed on-axis. We performed the reduction using version \texttt{v22.1}\footnote{\url{https://www.cosmos.esa.int/web/xmm-newton/sas-release-notes-2210}} of the \texttt{Science Analysis System} (\texttt{SAS}) including the data from the three European Photon Imaging Cameras (EPIC), i.e., MOS1, MOS2 and pn, on board XMM--Newton. We performed the standard data reduction method to obtain MOS and pn cleaned datasets, corresponding to 17\,ks for each instrument for the first observation after filtering out contamination from instrumental and background effects. For the second observation, the net exposures become 36\,ks and 32\,ks for MOS and pn, respectively.

NuSTAR observed PWN~G0.9+0.1 in 2021 (Obs ID: 40660007002, see Tab.~\ref{tab:xray_obs}). We processed the data with \texttt{NUSTARDAS v2.1.4} included in \texttt{HEASoft v6.35} using the NuSTAR calibration database (\texttt{CALDB}) version \texttt{20250331}\footnote{\url{https://heasarc.gsfc.nasa.gov/docs/heasarc/caldb/nustar/docs/release_20250331.txt}}. We used \texttt{nupipeline} to calibrate and clean the data. The net NuSTAR exposure times were 54\,ks and 53\,ks for the FPMA and FPMB modules, respectively.

Finally, we also analyzed one Chandra observation from 2000 (Obs ID: 1036, see Tab.~\ref{tab:xray_obs}) and processed it using \texttt{CIAO 4.17} with \texttt{CALDB} version \texttt{4.12.0}\footnote{\url{https://cxc.cfa.harvard.edu/caldb/downloads/Release_notes/CALDB_v4.12.0.html}}. The total Chandra exposure time after cleaning was 35\,ks. In this work, we reduce the Chandra data set with the sole purpose of obtaining a high-resolution image of the PWN. The soft X-ray spectral analysis is already well covered by the two XMM--Newton observations with larger photon statistics.

\begin{table}[ht]
    \centering
    \caption{Summary of the X-ray observations.}
    \label{tab:xray_obs}
    \resizebox{\linewidth}{!}{%
    \begin{tabular}{cccc}
        \toprule
        \toprule
        \multirow{2}{*}{Observatory} & \multirow{2}{*}{ObsID} & Date & Exposure \\
         &  &  (UT) & (ks) \\
        \midrule
        XMM-Newton & 0112970201 & 2000-09-23 &  18  \\
        
        XMM-Newton & 0144220101 & 2003-03-12 &  52  \\
        \midrule
        Chandra & 1036 & 2000-10-27 &  35 \\
        \midrule
        NuSTAR & 40660007002 & 2021-04-26  &  54 \\
        \bottomrule
    \end{tabular}%
    }
    \tablecomments{The exposure times are reported before the filtering process.}
\end{table}

\subsection{Spatial Analysis} \label{subsec:image_analysis}
\subsubsection{Chandra Spatial Analysis} \label{subsubsec:chandra_image}
\begin{figure}
    \centering
    \includegraphics[width=1.0\linewidth]{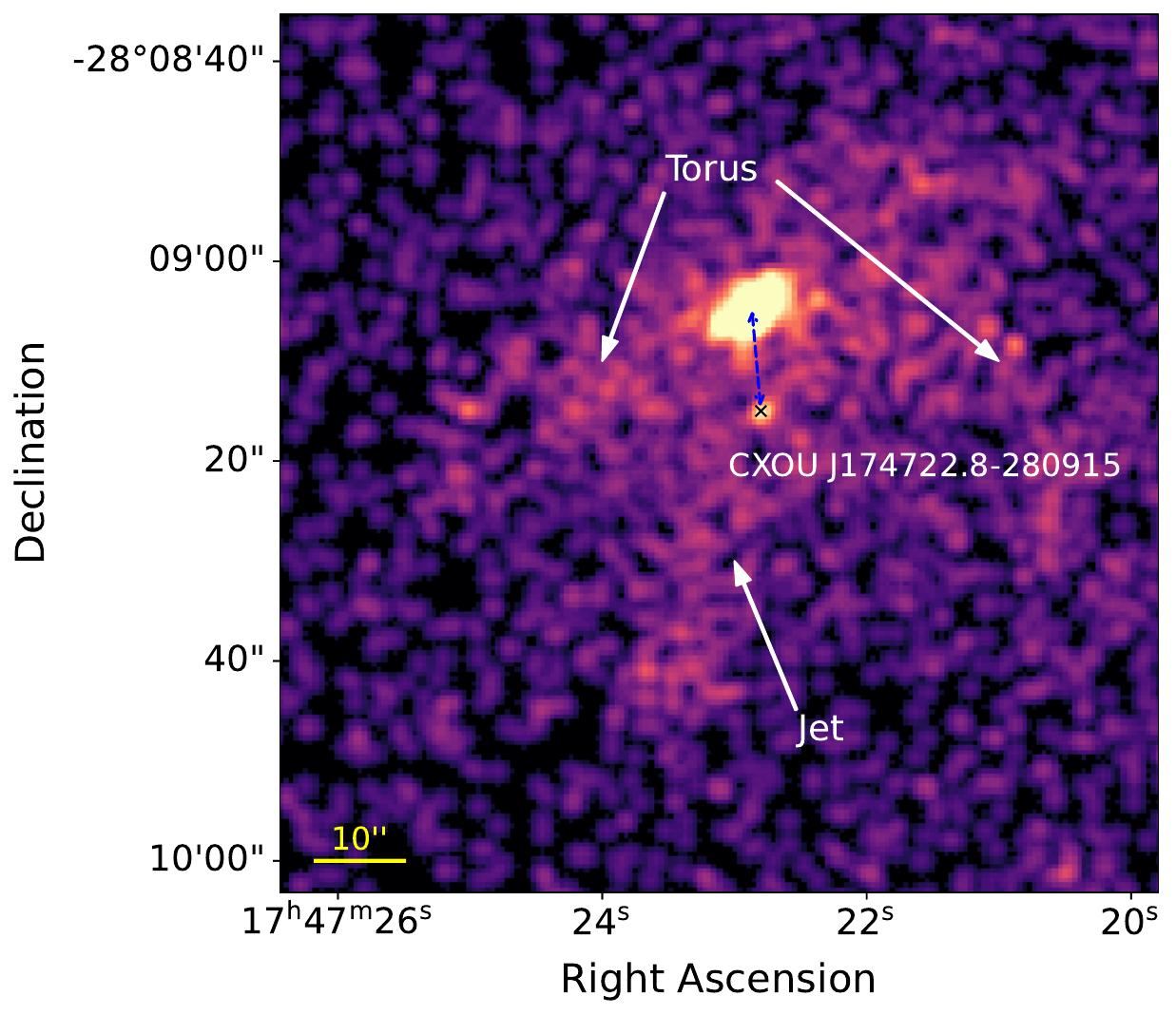}
    \caption{Exposure-corrected Chandra ACIS-S1 image of PWN~G0.9+0.1 in the energy range 2--7\,keV. We adopted an image bin size of one, corresponding to a pixel size of $0.492''$, and smoothed using a Gaussian kernel with radius $r=3$ and width $\sigma = 1.5$ pixels ($r=1.48''$, $\sigma=0.74''$). Solid arrows point to the torus (arc) and jet substructures of the PWN. A dashed arrow highlights the offset between the position of CXOU~J1747422.8--280915, marked with a cross, and the bright PWN core. CXOU~J1747422.8--280915 is most likely the X-ray counterpart to the pulsar.}
    \label{fig:chandra_image}
\end{figure}

Chandra was designed to achieve an excellent angular resolution of $0.5''$. This makes it the best X-ray telescope to study the complex morphology of PWNe. We produced the exposure-corrected image in the energy range 2--7\,keV using \texttt{fluximage} in \texttt{CIAO} from an exposure map generated using a median energy of 4.5\,keV. We applied to the image a bin size of 1 (pixel size $0.492''$), and smoothed with a Gaussian kernel of radius $r=3$ and Gaussian width $\sigma = 1.5$ pixels. The result is shown in Fig.~\ref{fig:chandra_image}. PWN~G0.9+0.1 displays a torus-jet morphology \citep[see also][]{Gaensler_2000} that is commonly observed in several young PWN systems. As shown in Fig.~\ref{fig:chandra_image}, the bright, compact $5'' \times 8''$ core of the PWN is centered $\sim10''$ away from the point source dubbed CXOU~J1747422.8--280915, the most likely X-ray counterpart of PSR~J1747--2809, even though no pulsations have been detected in the X-ray band. The semicircular arc, corresponding to the torus, has a radius of $\sim30''$, while the jet is extended by $\sim40''$ from the bright elliptical central clump towards the southeast. The overall PWN, considering also its substructure, can be encompassed in a region with a $\sim40''$ radius.

\subsubsection{XMM--Newton Spatial Analysis} \label{subsubsec:xmm_image}
\begin{figure}[ht]
    \centering
    \begin{subfigure}[b]{0.46\textwidth}
         \centering
         \includegraphics[width=\textwidth]{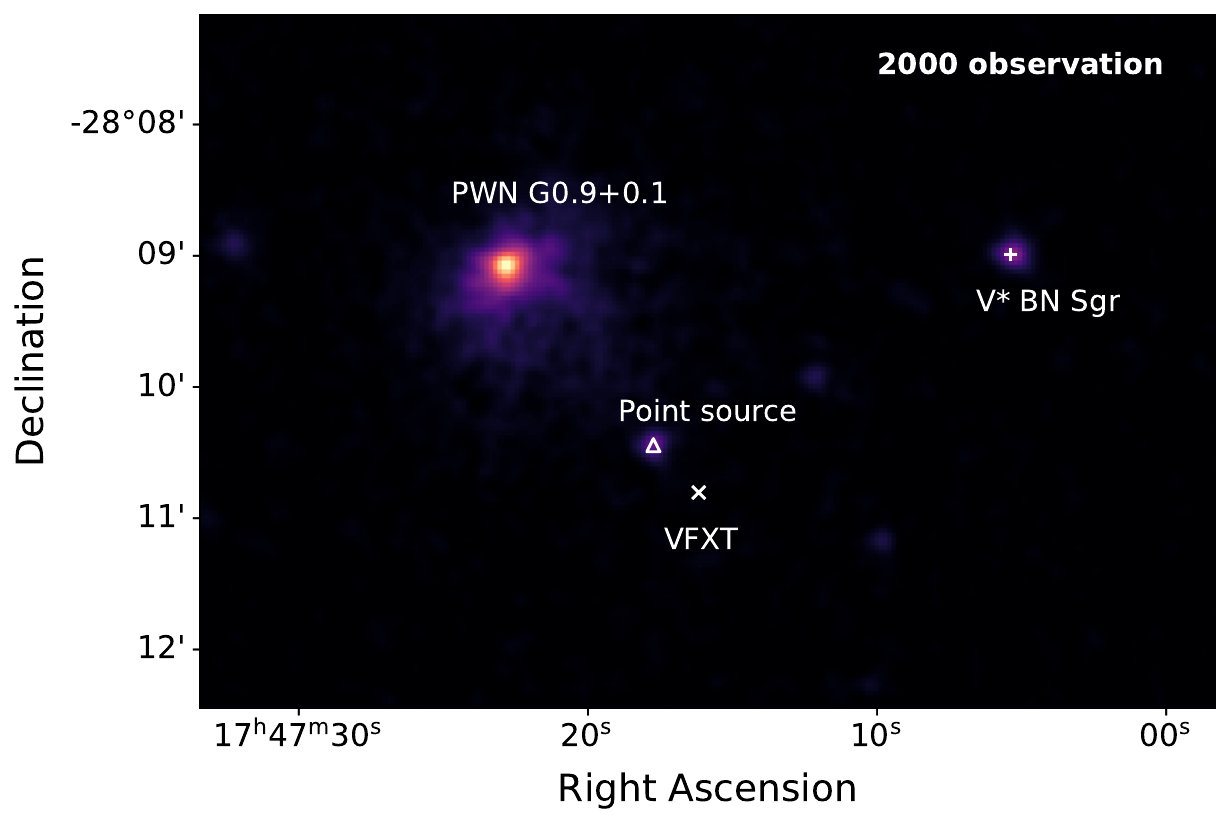}
         \caption{}
         \label{subfig:xmm_2000}
     \end{subfigure}
     \begin{subfigure}[b]{0.46\textwidth}
         \centering
         \includegraphics[width=\textwidth]{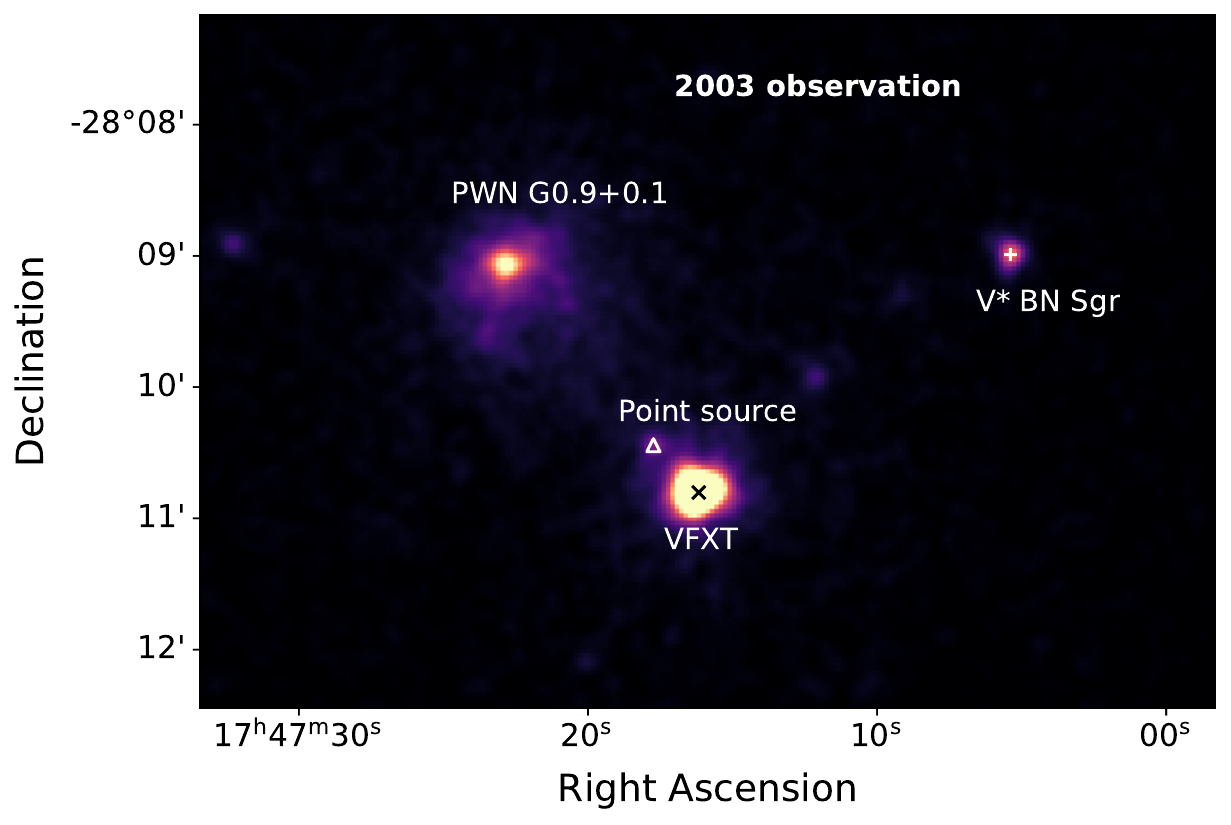}
         \caption{}
         \label{subfig:xmm_2003}
     \end{subfigure}
    \caption{(a) Combined EPIC-MOS and EPIC-pn image of PWN~G0.9+0.1 and its surrounding region for the 2000 observation. (b) Combined XMM EPIC-MOS image of PWN~G0.9+0.1 and its surrounding region for the 2003 observation. In both images, the energy range is set to 2--10\,keV, and we use a Gaussian smoothing kernel with radius $r=3$ and $\sigma = 1.5$ pixels. We marked with crosses the positions of VFXT XMMU~J174716.1--281048 and V* BN Sgr, and with a triangle the position of the point source 4XMM~J174717.7--281026.}
    \label{fig:XMM_image}
\end{figure}

While Chandra allowed us to study the morphology of PWN~G0.9+0.1, XMM--Newton can give us a better view of the region surrounding the source due to its large FoV and better sensitivity. We produced the XMM--Newton images for both observations to study PWN~G0.9+0.1 and its surroundings in the two epochs. We obtained the images for all EPIC instruments of each observation using the \texttt{SAS} task \texttt{eimageget} in the 2--10\,keV energy range. We combined the results obtained for the MOS and pn cameras using \texttt{eimagecombine} to derive Fig.~\ref{fig:XMM_image}. For the 2003 data set, we were able to combine only the MOS1 and MOS2 images since the pn data were taken in Large Window mode, which is not supported by \texttt{eimageget} and \texttt{eimagecombine}. In the final combined images from each observation shown in Fig.~\ref{fig:XMM_image}, there are nearby sources to PWN~G0.9+0.1. The first source, detected in both observations and labeled HD~161507, is classified as a star that is part of the Algol-type eclipsing binary V* BN Sgr \citep{Budding_2004}. The second source is serendipitously detected in the second observation and classified as a very-faint X-ray transient (VFXT) dubbed XMMU~J174716.1--281048 \citep{Sidoli_2003_ATel}. We present a more detailed analysis of this transient source in Sect.~\ref{sec:xmmu}. There is also a point source close to the VFXT in both observations, and it is reported in the 4XMM-DR14 catalog as 4XMM~J174717.7--281026. 

\begin{figure*}[ht]
    \centering
    \includegraphics[width=\linewidth]{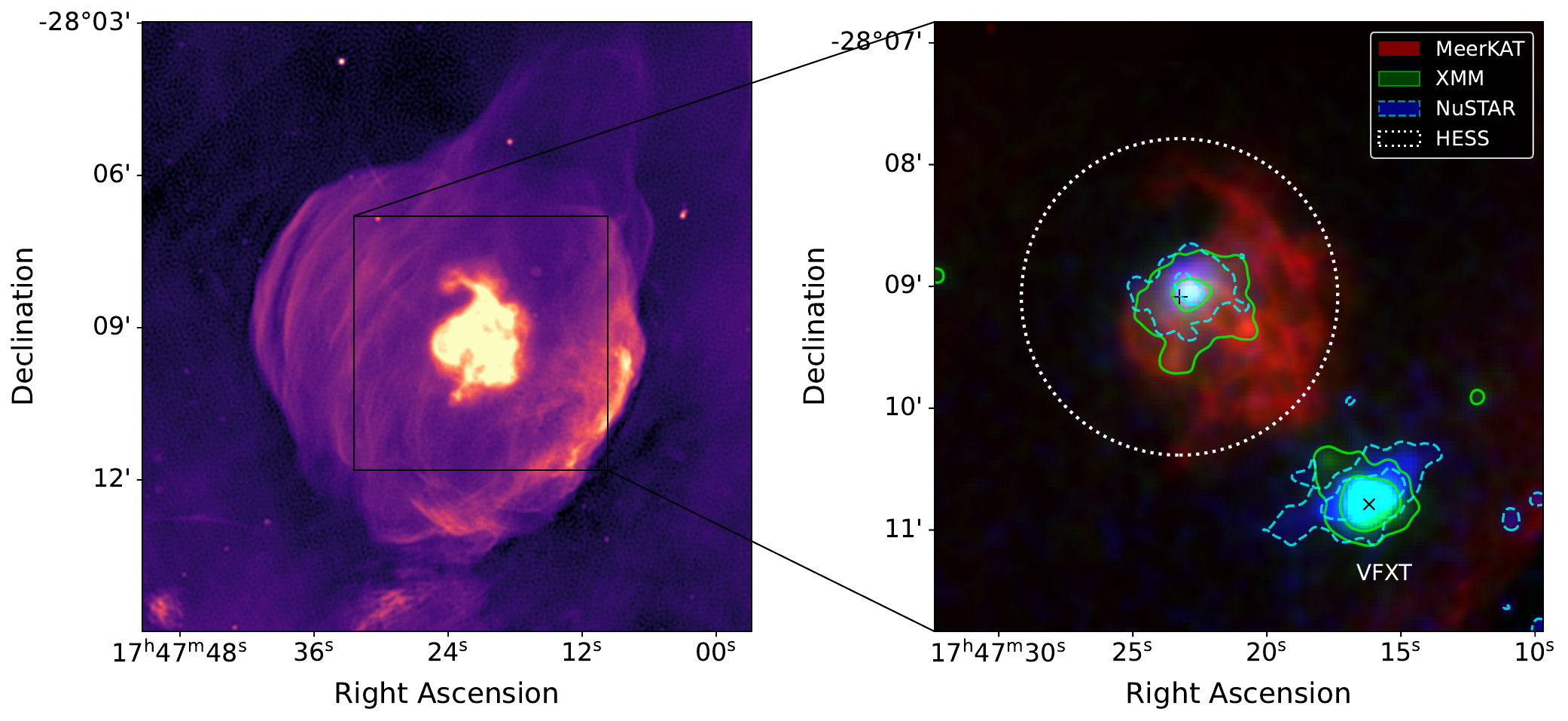}
    \caption{\textit{Left panel:} MeerKAT 1.28\,GHz image \citep{MeerKAT_2022} of the SNR shell associated with PWN~G0.9+0.1. \textit{Right panel:} RGB image of PWN~G0.9+0.1, including observations from MeerKAT at 1.28 GHz \citep[red, from][]{MeerKAT_2022}, XMM--Newton at 2--10\,keV (green, this work), and NuSTAR at 10--30\,keV (blue, this work). We also show the 95\% and 99\% confidence contours levels of XMM (solid) and NuSTAR (dashed). The best-fit position and 95\% confidence level upper limit on the source extension as observed by HESS \citep{HESS_2005} are shown as the plus marker and dotted circle. The position of the VFXT XMMU~J174716.1--281048 is marked with a black cross.}
    \label{fig:multiwaveImage_withSNR}
\end{figure*}

In Fig.~\ref{fig:multiwaveImage_withSNR} we show a zoom-in of the MeerKAT 1.28\,GHz GC mosaic of the GC \citep{MeerKAT_2022} of the SNR shell and the PWN. The XMM--Newton 2--10\,keV image of the 2003 observation is represented in green color and contours in the RGB image of the PWN on the right panel.

\subsubsection{NuSTAR Spatial Analysis} \label{subsubsec:nustar_image_analysis}
\begin{figure*}[ht]
    \centering
    \begin{subfigure}{0.3\textwidth}
         \centering
         \includegraphics[width=\textwidth]{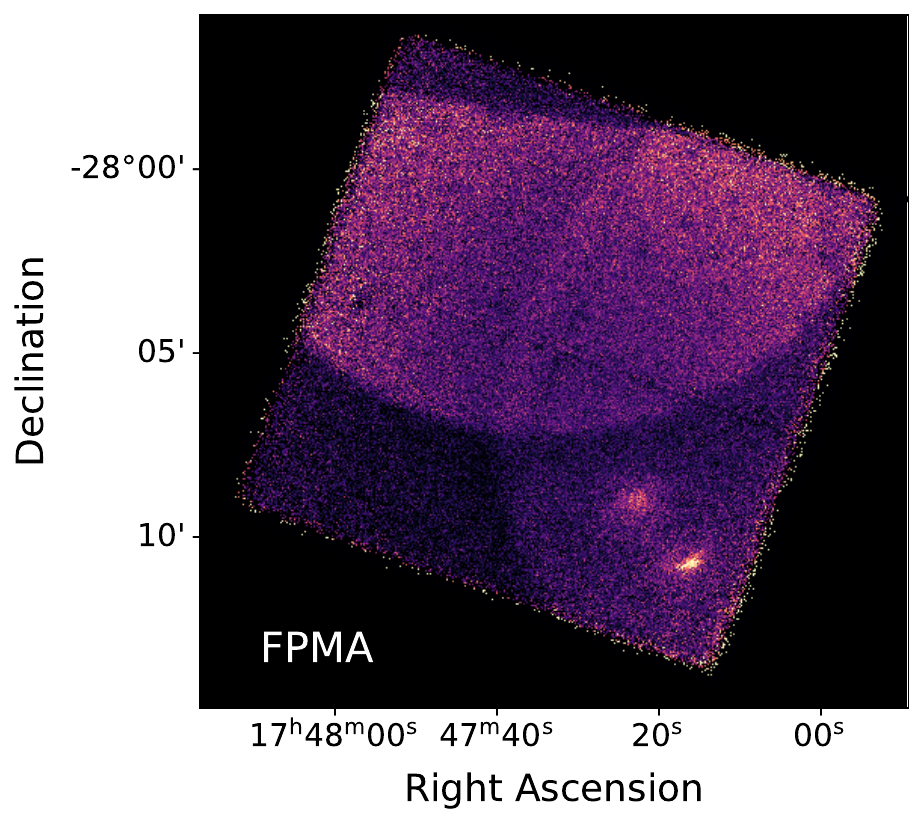}
         \caption{}
         \label{subfig:fpma_fov}
     \end{subfigure}
     \begin{subfigure}{0.3\textwidth}
         \centering
         \includegraphics[width=\textwidth]{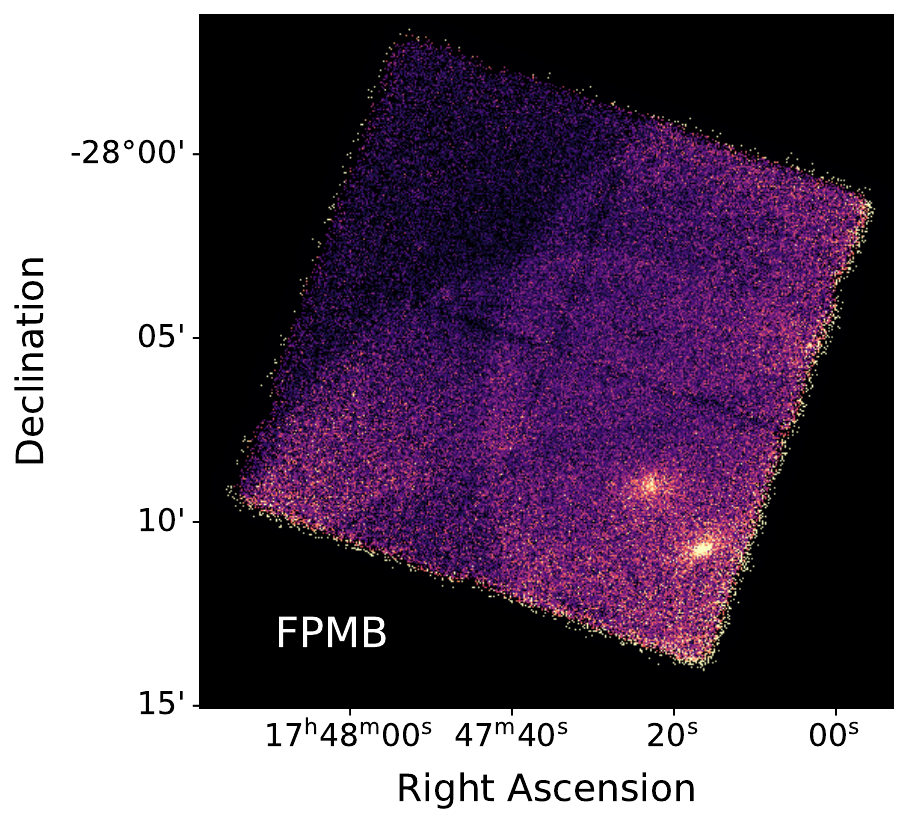}
         \caption{}
         \label{subfig:fpmb_fov}
     \end{subfigure}
     
    \begin{subfigure}{0.3\textwidth}
         \centering
         \includegraphics[width=\textwidth]{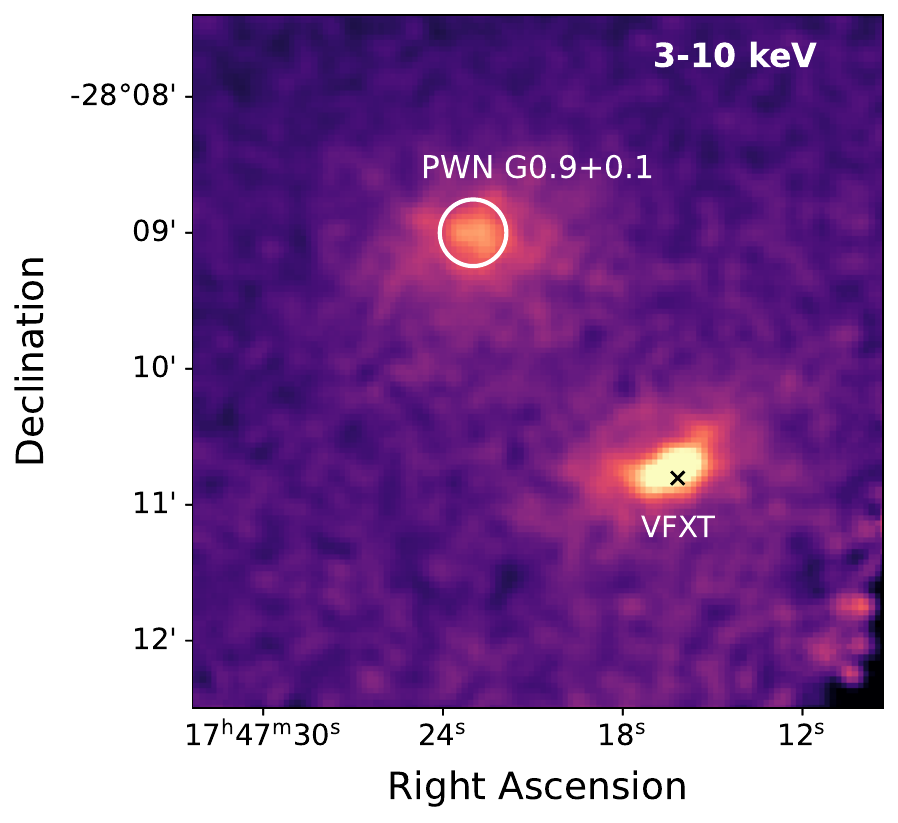}
         \caption{}
         \label{subfig:nustar_soft}
     \end{subfigure}
     \begin{subfigure}{0.3\textwidth}
         \centering
         \includegraphics[width=\textwidth]{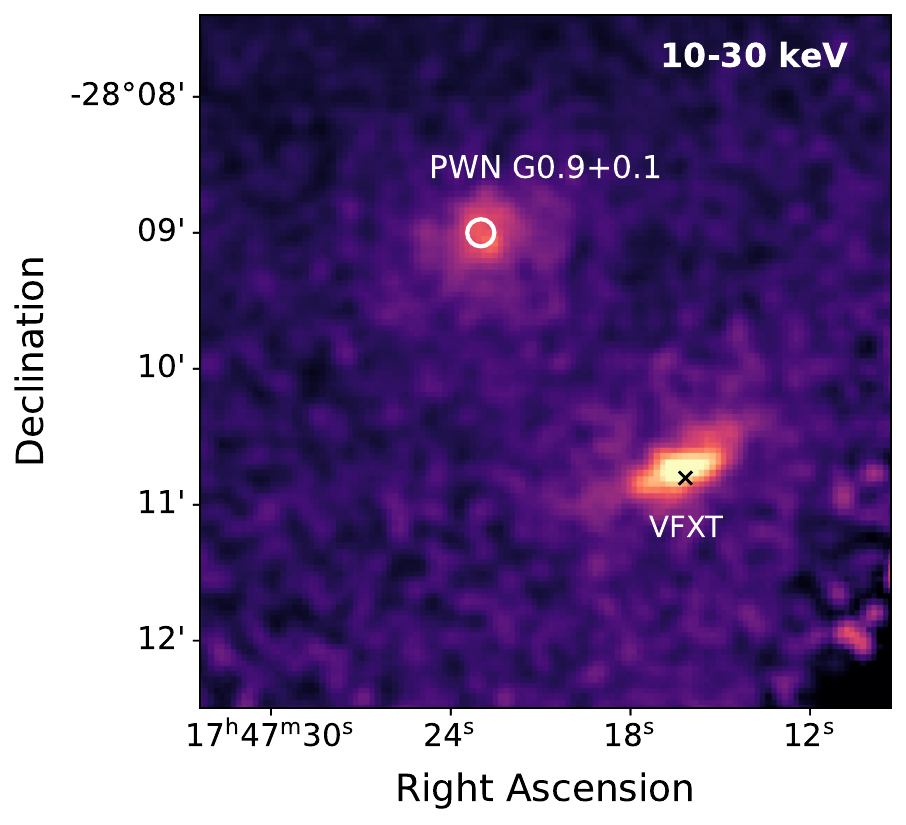}
         \caption{}
         \label{subfig:nustar_hard}
     \end{subfigure}
     \begin{subfigure}{0.38\textwidth}
         \centering
         \includegraphics[width=\textwidth]{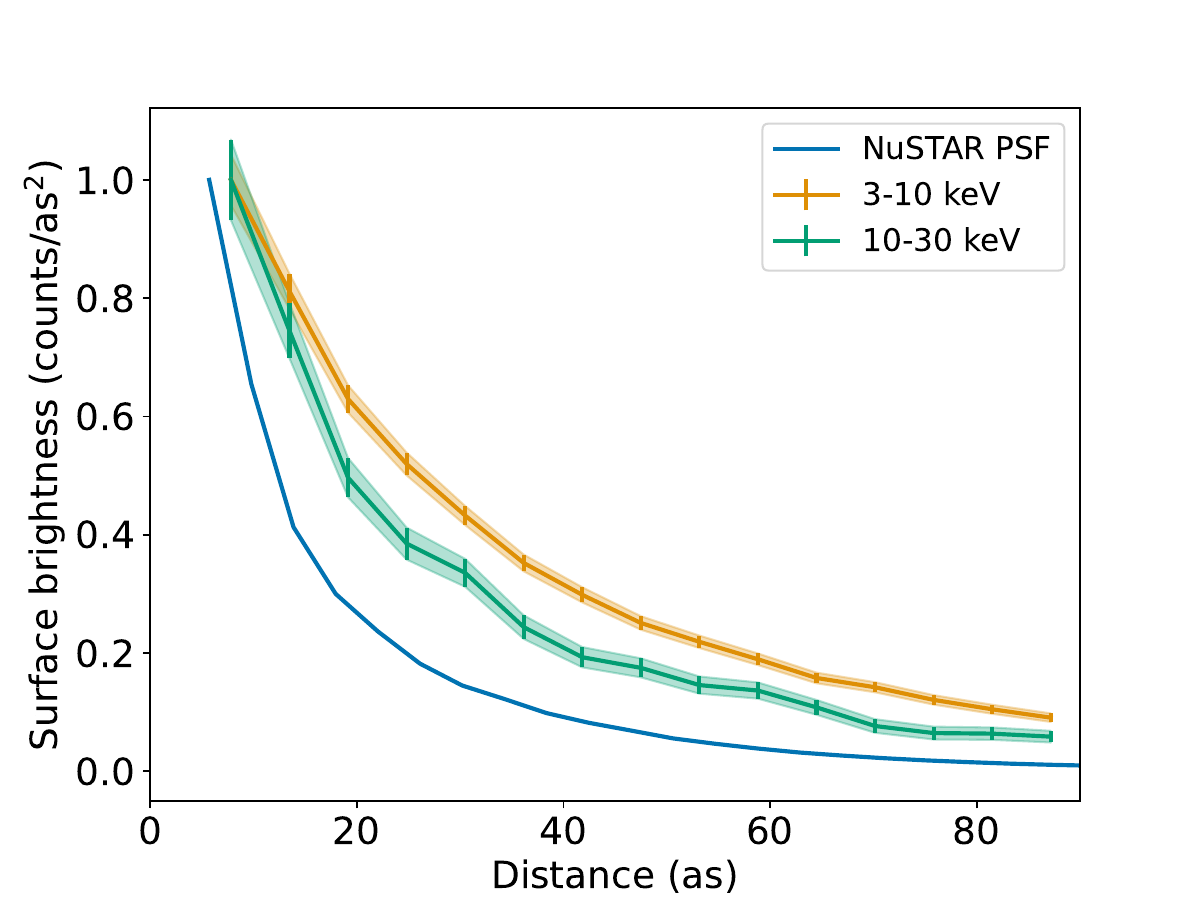}
         \caption{}
         \label{subfig:nustar_radProf}
     \end{subfigure}  
     \caption{FPMA (a) and FPMB (b) images of the NuSTAR FoV in the 3--79\,keV energy range, highlighting the stray light pattern affecting the observation. (c) Merged (FPMA and FPMB), exposure- and vignetting-corrected NuSTAR image of PWN~G0.9+0.1 in the 3--10\,keV energy range, smoothed with a Gaussian kernel with $r=3$ and $\sigma = 1.5$ pixels. The solid circle represents the radial extension of PWN~G0.9+0.1 estimated with \texttt{Sherpa}. The cross marks the position of VFXT XMMU~J174716.1--28104. (d) Same as for (c), but in the 10--30\,keV range. (e) Normalized (to one) radial profiles of PWN~G0.9+0.1 in the 3--10\,keV (orange) and 10--30\,keV (green) energy ranges, compared to the PSF of NuSTAR (blue solid curve). The shaded areas represent the $1\sigma$ uncertainty bands.}
    \label{fig:nustar_imageAnalysis}
\end{figure*}

Thanks to the extended energy range of NuSTAR when compared to XMM--Newton or Chandra, we can image PWN~G0.9+0.1 for the first time above 10\,keV. We produced the exposure-corrected image of the NuSTAR FoV in the full energy range of the telescope (3--79\,keV), for both FPMA and FPMB instruments, using the exposure map produced with \texttt{nupipeline}. The results are shown in Fig.~\ref{subfig:fpma_fov} and Fig.~\ref{subfig:fpmb_fov} for FPMA and FPMB, respectively. Both images illustrate the strong contamination from stray light effects (see Section~\ref{subsubsec:nustar_spec_analysis} for details on background treatment considering the stray light contamination). In addition to the detection of PWN~G0.9+0.1, NuSTAR also detected emission coincident in location with the VFXT XMMU~J174716.1--281048. The VFXT is also clearly visible in the 10--30\,keV band and is the very first detection of this X-ray transient above 10\,keV. More details on its spectrum as observed by NuSTAR are discussed in Sect.~\ref{sec:xmmu}.

We obtained images for NuSTAR FPMA and FPMB in the 3--10\,keV and 10--30\,keV energy ranges using the \texttt{HEASoft} task \texttt{extractor}. We used the \texttt{nuexpomap} task of \texttt{NuSTARDAS} to obtain the corresponding exposure maps, applying the vignetting correction. The vignetting effect is energy dependent, so we chose a median energy for each of the analyzed bands: 6.5\,keV for the first and 20\,keV for the second. Finally, FPMA and FPMB images are combined with their exposure maps using \texttt{XIMAGE} to obtain Fig.~\ref{subfig:nustar_soft} and Fig.~\ref{subfig:nustar_hard} for the soft (3--10\,keV) and hard (10--30\,keV) energy bands, respectively. The NuSTAR 10--30\,keV image is also represented in blue in the RGB image of the PWN on the right panel of Fig.~\ref{fig:multiwaveImage_withSNR}. The size of PWN~G0.9+0.1 decreases with increasing energy, hinting at the presence of the synchrotron burnoff effect of the highest energy electrons. To better quantify this effect, we produced the radial profiles of the source in annuli with radii between $5''$ and $90''$ for the two energy bands and compared them to the point spread function (PSF) of NuSTAR. The results are displayed in Fig.~\ref{subfig:nustar_radProf}. The source emission is clearly extended in both energy bands, and the narrow profile of the higher energy band shows a clear indication of the synchrotron burnoff effect. We then measured the source size from the exposure- and vignetting-corrected images using \texttt{Sherpa} in \texttt{CIAO}. We fitted a PSF-convolved 2D Gaussian model, including a constant that represents the background model component. We obtained a full width at half maximum of $14''.7 \pm 1''.8$ for the 3--10\,keV energy band and ($6'' \pm 2''$) for the 10--30\,keV energy band, respectively.

\subsection{Spectral Analysis} \label{subsec:spec_analysis}
\subsubsection{XMM-Newton Spectral Analysis} \label{subsubsec:xmm_spec_analysis}
We analyzed the 2000 and 2003 observations, including data from all three EPIC cameras onboard XMM--Newton. We extracted the spectra of PWN~G0.9+0.1 from a circular $r=45''$ region centered on the X-ray peak in the XMM--Newton images, motivated by the X-ray size estimated from the Chandra image ($\sim40''$). All the spectra were grouped with a minimum of 25 counts per bin. The background spectra were extracted from an $80''$ circular region selected in the same chip of each CCD of each observation. The selected regions were free of bright sources, chip gaps, and CCD edges. The choice of the background region did not impact the final results, as we tested different background regions, both in location and size, and obtained consistent results within errors. We simultaneously fitted the spectra of the three cameras in the energy range 2--10\,keV using an absorbed power law (\texttt{tbabs*powerlaw}), assuming \texttt{wilm} abundance table for the absorbing column density \citep{Wilms_2000}, and choosing the pn as the reference spectrum since it provides the best signal to noise in both observations. We first performed the analysis of the two epochs separately and obtained compatible results within the statistical uncertainties. A more detailed description is included in Appendix~\ref{app:xmm_exta}.

Next, we simultaneously fitted the two epochs, fixing the reference spectrum to that of the image with the best statistics, the 2003 pn observation. We used a calibration constant for the spectral models from the 2003 MOS, 2000 pn, and 2000 MOS images. We obtained a reduced $\chi^2$ value for 545 degrees of freedom (dof) of $\chi^2/dof=0.87$ for a photon index of $\Gamma=1.83 \pm 0.11$ and a column density of $N_H=1.96^{+0.13}_{-0.11}\times 10^{23}\ \text{cm}^{-2}$. Extending the energy range down to 0.5\,keV does not improve the error bars on the column density. The unabsorbed 3--10\,keV flux is $F_{3-10}= 3.4^{+0.1}_{-0.2}\times 10^{-12}\ \text{erg}^{-2}\ \text{s}^{-1}\ \text{cm}^{-2}$. The result is reported in the first row of Tab.~\ref{tab:PL_fit}. No other component is required in the model, such as one that describes a pulsar component. In particular, the pulsar X-ray luminosity is $\le 1$\% of the PWN X-ray luminosity, as estimated by \cite{Gaensler_2000}. The non-detection of pulsed X-ray emission could be due to a combination of an intrinsically soft spectrum and the high measured column density, and neglecting its contribution in the PWN spectral analysis does not affect the final result.

To study the presence of photon index variations with distance from the pulsar, we focused on the observation with the deepest exposure (2003 epoch). We selected three regions, chosen to each have roughly the same number of source counts: an inner $r < 15''$ circle and two concentric annuli with radii $15'' < r < 27.5''$ and $27.5'' < r < 40''$. We extract their spectra from each EPIC camera as before and perform a fit simultaneously on all three. We tested for the pulsar spectral contribution in the innermost region and found no evidence for an additional component in the spectrum. From this, we measured the radial profile of the photon index and the surface brightness. The photon index is observed to increase from $\Gamma = 1.4 \pm 0.1$ in the innermost region to $\Gamma = 2.09 \pm 0.14$ in the outermost region, a spectral steepening also reported by \cite{Porquet_2003} and \cite{Holler_2012}. However, the hard photon index observed in the innermost region ($\Gamma \sim 1.4$), along with the high estimated column density ($N_H \sim 10^{23}\ \text{cm}^{-2}$), might be an indication of contamination by dust scattering. It is important to take into account this effect when studying highly-absorbed sources, as the scattering of X-ray photons generates diffuse halos around the affected sources and influences their observed photon index and flux, making them appear harder and fainter \citep{Predehl_1995, Smith_2016, Jin_2017, Jin_2018, Jin_2019}. We explored the effects of scattering in Section~\ref{subsubsec:dust}

\subsubsection{NuSTAR Spectral Analysis} \label{subsubsec:nustar_spec_analysis}
Due to stray light contamination (see Fig.~\ref{subfig:fpma_fov} and \ref{subfig:fpmb_fov}) and to the presence of the nearby VFXT source XMMU~J174716.1--281048, the choice of both the source and background regions is difficult. We explore the most accurate source spectral region from the NuSTAR observation, considering the XMM--Newton results while minimizing both background and VFXT contamination. We found that a radius of $50''$ is the most reasonable choice. We extracted the spectrum from FPMA and FPMB in the energy range 3--30\,keV using an absorbed power law model, as was done for XMM--Newton. We obtained $\chi^2/dof=0.85$ for $N_H=(1.9 \pm 0.4) \times 10^{23}\ \text{cm}^2$ and $\Gamma=2.20^{+0.13}_{-0.12}$, reported in the second row of Tab.~\ref{tab:PL_fit}. XMM--Newton finds a harder spectrum ($\Gamma_{\text{XMM}} \sim 1.8$ in Tab~\ref{tab:PL_fit}), though both instruments find compatible unabsorbed fluxes in the overlapping 3--10\,keV energy range ($F_{\rm XMM}= 3.4^{+0.1}_{-0.2}\times 10^{-12}\ \text{erg}^{-2}\ \text{s}^{-1}\ \text{cm}^{-2}$, $F_{\rm NuSTAR}= 3.2^{+0.4}_{-0.3}\times 10^{-12}\ \text{erg}^{-2}\ \text{s}^{-1}\ \text{cm}^{-2}$). The discrepancy may be evidence for a spectral break, or cutoff, in the PWN X-ray spectrum, or may be due to the differences in the instrument performance and calibration techniques. We investigated these scenarios by performing a joint fit of the two instruments, see Sect.~\ref{subsubsec:jointFit}. Another possible explanation is the influence of dust scattering, which was studied in Sect.~\ref{subsubsec:dust}.

We carried out a spatially-resolved X-ray analysis following the methods applied for the XMM--Newton data set (see Sect.~\ref{subsubsec:xmm_spec_analysis}) using three regions: $r \lesssim 24''$, $24'' \lesssim r \lesssim 38''$, and $38'' \lesssim r \lesssim 50''$. We simultaneously fitted the FPMA and FPMB spectra and derived the radial profiles for the photon index and surface brightness. We observe almost a constant trend in the photon index, which changes from $\Gamma = 2.00^{+0.09}_{-0.10}$ in the innermost region to $\Gamma = 2.27 \pm 0.17$ in the outermost, with no variation between the two annuli. Even though dust scattering is less prominent in the NuSTAR energy band, as the flux and size of the dust scattering halo decrease with energy, we revised the radial profiles considering this effect (see Sect.~\ref{subsubsec:dust}).

\begin{table*}[ht]
    \centering
    \caption{Summary of the results for the absorbed power law model fit for XMM--Newton, NuSTAR, and the joint XMM--Newton + NuSTAR fits.}
    \label{tab:PL_fit}
    \begin{tabular}{lcccccc}
        \toprule
        \toprule
        \multirow{2}{*}{Data set} & Energy & $N_H$ & \multirow{2}{*}{$\Gamma$} & $F_{3-10}$ & \multirow{2}{*}{$\chi^2/dof$} & \multirow{2}{*}{Constant} \\
         & (keV) & ($10^{23}\ \text{cm}^{-2}$) &  & ($10^{-12}\ \text{erg}\ \text{s}^{-1}\ \text{cm}^{-2}$) &  &  \\
        \midrule
        \multirow{3}{*}{XMM} & \multirow{3}{*}{2--10} & \multirow{3}{*}{$1.96^{+0.13}_{-0.11}$} & \multirow{3}{*}{$1.83 \pm 0.11$} & \multirow{3}{*}{$3.4^{+ 0.1}_{-0.2}$} & \multirow{3}{*}{472.3/545} & $0.96 \pm 0.04$ (M03) \\
         &  &  &  &  &  & $0.92 \pm 0.04$ (P00)\\
         &  &  &  &  &  & $0.88 \pm 0.04$ (M00)\\
         \midrule
        NuSTAR & 3--30 & $1.9 \pm 0.4$ & $2.20^{+0.13}_{-0.12}$ & $3.2^{+0.4}_{-0.3}$ & 256.5/303 & $1.06 \pm 0.06$ (B) \\
        \midrule
        \multirow{4}{*}{XMM + NuSTAR} & \multirow{4}{*}{2--30} & \multirow{4}{*}{$2.19 \pm 0.09$} & \multirow{4}{*}{$2.12 \pm 0.07$} & \multirow{4}{*}{$3.6^{+0.1}_{-0.2}$} & \multirow{4}{*}{762.0/851} & $0.96 \pm 0.04$ (M03) \\
         &  &  &  &  &  & $0.92 \pm 0.05$ (P00)\\
         &  &  &  &  &  & $0.88 \pm 0.04$ (M00)\\
         &  &  &  &  &  & $0.92 \pm 0.04$ (N)\\
        \bottomrule
    \end{tabular}
    \tablecomments{The cross-calibration constants are labeled according to the instrument they are referring to: ``M" is MOS, ``P" is pn, ``N" is NuSTAR (FPMA and FPMB), and ``B" is FPMB. ``00" refers to the 2000 XMM--Newton epoch, ``03" to the 2003 epoch. All uncertainties are reported at the 90\% confidence level.}
\end{table*}

\subsubsection{Joint XMM--Newton and NuSTAR Spectral Fit} \label{subsubsec:jointFit}
To investigate the discrepancy in the photon index observed between the XMM--Newton and NuSTAR results, we jointly fitted the two data sets. This method has been extensively used for X-ray PWNe, see for example \cite{Bamba_2022}, and it allows us to combine the soft band (derived from either XMM--Newton, Chandra, or Suzaku) with the hard band of NuSTAR to study the X-ray spectrum in a broad energy range. It is particularly useful to understand whether cutoffs or breaks in the spectrum are observed.

\begin{figure}[ht]
    \centering
    \includegraphics[width=1.0\linewidth]{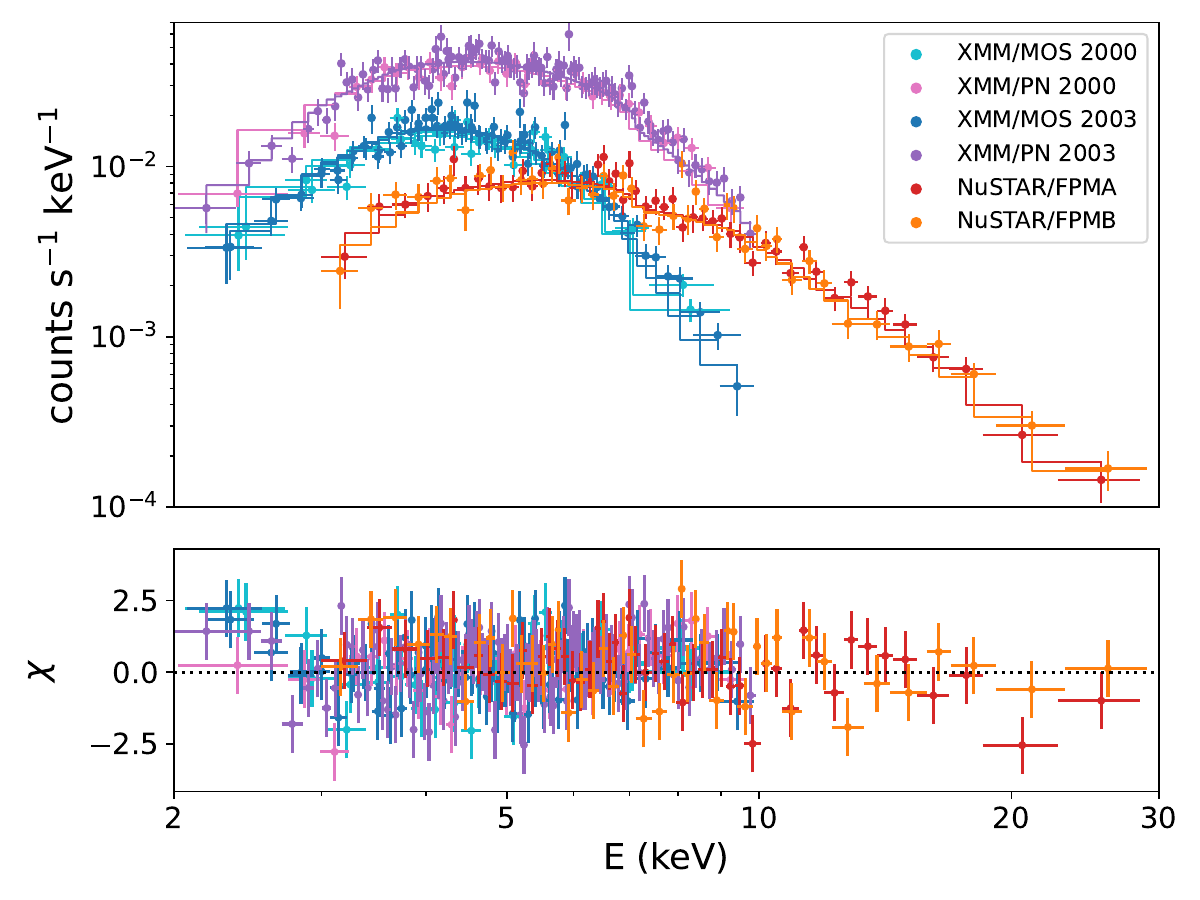}
    \caption{Joint fit of both XMM--Newton observations together with the new NuSTAR data of PWN~G0.9+0.1 assuming an absorbed power law model in the energy range 2--30\,keV (upper panel), along with the model residuals (bottom panel) expressed as (data - model)/error. The joint fit parameters are reported in the last row of Tab.~\ref{tab:PL_fit}.}
    \label{fig:jointFit}
\end{figure}

The joint fit is performed using XMM--Newton data in the 2--10\,keV energy range and NuSTAR data in the 3--30\,keV energy range. We adopted a constant linking FPMA and FPMB NuSTAR spectra and a constant linking the two MOS spectra in each epoch. We performed the fits individually for each XMM--Newton observation and also for both XMM--Newton observations and NuSTAR data together. The results for the two epochs agree within statistical uncertainties, as described in Appendix~\ref{app:xmm_exta}. For the joint fit of all three data sets, we fixed the reference spectrum to that of the 2003 pn observation. We obtained $\chi^2/dof=0.9$ for a photon index $\Gamma=2.11 \pm 0.07$ and a column density of $N_H=(2.19 \pm 0.09) \times 10^{23}\ \text{cm}^2$. The 3--10\,keV flux is $F_{\rm 3-10 \ keV}= 3.6^{+0.1}_{-0.2}\times 10^{-12}\ \text{erg}^{-2}\ \text{s}^{-1}\ \text{cm}^{-2}$. We display the joint fit from all three observations in Fig.~\ref{fig:jointFit} and report the result in the last row of Tab~\ref{tab:PL_fit}. Without the constants in the joint fit model, we still obtain a reasonable fit, $\chi^2/dof = 0.92$ for $N_H=22.2 \pm 0.9$, $\Gamma=2.15 \pm 0.06$, and a compatible 3--10\,keV flux ($F_{\rm 3-10\ keV}= (3.4 \pm 0.1)\times 10^{-12}\ \text{erg}^{-2}\ \text{s}^{-1}\ \text{cm}^{-2}$).

Finally, we also investigated the presence of possible features, such as a break or cutoff, in the spectrum following the approach presented in \cite{Bamba_2022}. We tested a broken power law and an exponential cutoff power law. In both cases, the fit returned a value $\chi^2/dof=0.87$, comparable with the simple power law model. In the case of the broken power law, we found a break energy of $E_{\rm break}=6.6^{+1.4}_{-0.7}$\,keV. For the cutoff power law, we found $E_{\rm cut}=15^{+10}_{-5}$\,keV. These two models can both achieve similar fit results as the simple power law does; thus, a spectral break or a cutoff could be possible, but they are not clearly present. Moreover, the break and cutoff energies we obtained are both close to $\sim10$\,keV, the upper boundary for the XMM--Newton data. This could imply that these features could be related to the transition from one instrument to the other, and not to the physics of the X-ray emission. Overall, a simple power law can reconcile the observed differences between the photon indices of XMM--Newton and NuSTAR.

\subsubsection{Dust Scattering Correction} \label{subsubsec:dust}
Dust grains in the interstellar medium affect the propagation of X-ray photons, and this implies that X-ray sources located behind sufficiently dense dust clouds are expected to be surrounded by extended and faint X-ray-emitting halos \citep{Predehl_1995}. The effect becomes more pronounced when the column density is high ($N_H \sim 10^{22-23}$\,cm$^{-2}$). The scattering cross section is larger for lower energy photons, therefore, dust scatters preferentially soft X-ray photons. This has an effect on the observed source spectrum, unless all the photons of the source and halo are accumulated in the spectral extraction region. Consequently, the spectrum of a source extracted from a region smaller than the full extent of the dust halo will appear fainter and harder than the intrinsic one. With the X-ray analysis we conducted and presented in Sect.~\ref{subsubsec:xmm_spec_analysis} and~\ref{subsubsec:nustar_spec_analysis}, we obtained $N_H \sim 2 \times 10^{23}$ cm$^{-2}$ for PWN~G0.9+0.1. Moreover, the spatially-resolved XMM--Newton analysis returned a relatively hard index ($\Gamma\sim 1.4$) in the innermost ($r<15''$) region. Both these conditions suggest the presence of significant contamination by dust scattering in the spatially resolved analysis results. As explained above, the effects of dust scattering become less prominent when selecting larger spectral extraction regions, and, for this reason, we did not take into account the phenomenon when studying the overall PWN spectrum.
 
To understand the impact of dust scattering, we used the model developed by \cite{Jin_2017}. It was initially deployed on AX~J1745.6--2901, an X-ray transient close ($d\sim 1.45'$) to the GC, and has since been applied to several other sources in the GC region \citep{Jin_2018, Jin_2019}. To model the scattering of X-ray photons around this transient, they assumed the dust to be organized in layers located at different distances on the line of sight between the observer and the source. To reproduce the radial profile of the AX~J1745.6--2901 halo, two dust layers are required. The first (``Layer 1") is located very close to the source (at a fractional distance lower than 0.1 from the transient) and accounts for the scattering component generated by the material close to the source and to the GC. The second (``Layer 2") is located closer to Earth and accounts for the component due to the material inside the Galactic disc. The latter includes the majority of dust ($\sim75$\%), and is estimated to be located at $\sim0-5$\,kpc from Earth, overlapping with some of the Milky Way's spiral arms. A visual representation of the geometry can be found in the figures of \cite{Jin_2017}. The effects of the two layers on the source spectrum can be included in \texttt{xspec} using the multiplicative tables \texttt{axjdust} (Layer 1) and \texttt{fcgdust} (Layer 2). The main advantage of this description is that both components incorporate the shape of the PSF for the corresponding instrument in the dust scattering model, allowing a more accurate assessment of the effect. Including the PSF in the treatment is relevant because scattering halos are often more extended than the instrument's PSF and thus not fully covered when selecting small spectral extraction regions.

To model the dust around PWN~G0.9+0.1, we assumed that the foreground component along the line of sight is the same as for AX~J1745.6--2901, since the PWN is close to the GC. We applied the \texttt{fcgdust} model to both XMM--Newton and NuSTAR data, to correct the photon index and surface brightness profile measurements, accounting for dust scattering. We additionally tested the \texttt{xspec} model \texttt{xscat} to model dust scattering and found slightly different results and larger error bars than from \texttt{fcgdust}, which is most likely due to the fact that the \texttt{xscat} model does not take into account the instrument PSF. The photon index and surface brightness for each region and instrument, before and after the correction, are shown in Tab.~\ref{tab:dustScatterning}. The corrected photon index radial profile for XMM--Newton shows spectral steepening where $\Gamma = 1.93 \pm 0.11$ in the innermost region and $\Gamma = 2.05 \pm 0.15$ in the outermost region. The NuSTAR photon index radial profile shows $\Gamma=2.16 \pm 0.10$ in the innermost region and $\Gamma = 2.29 \pm 0.17$ in the outermost region. As we would expect, the difference is less prominent in the NuSTAR data, since the energy range (3--79\,keV) avoids most scattering effects and the extraction regions are larger than those selected for XMM--Newton.

We note that the photon indices we obtain for XMM--Newton are systematically harder than those obtained with NuSTAR, even after the dust scattering correction. The differences could be due to an intrinsically curved spectrum or to imperfections in the modeling of the dust layers. We observe a photon index variation in the innermost XMM--Newton region from $\Gamma_{\rm before}=1.44 \pm 0.11$ to $\Gamma_{\rm after}=1.93 \pm 0.11$ ($\Delta \Gamma \sim 0.5$) which is much larger than the difference observed between the XMM--Newton ($\Gamma_{\rm XMM}=1.93 \pm 0.11$) and the NuSTAR ($\Gamma_{\rm NuSTAR}=2.16 \pm 0.10$) index. This supports the hypothesis that the systematic difference between XMM--Newton and NuSTAR is likely due to the simple assumption made on the dust scattering layer, i.e., that it is the same as for AX~J1745.6--2901. Furthermore, we tested the NuSTAR spectrum in the same energy range as XMM--Newton and found a photon index of $\Gamma=2.2 \pm 0.3$, which is compatible with the index $\Gamma=1.83 \pm 0.11$ for the XMM--Newton analysis, reported in Tab.~\ref{tab:PL_fit}. The large uncertainties obtained on this test do not allow us to exclude either possibility with the current samples. Deeper XMM--Newton and NuSTAR observations, as well as a specific treatment of the dust scattering layers on the line of sight to PWN~G0.9+0.1, could improve the results in the future.

\begin{table*}[ht]
    \centering
    \caption{Summary of the results for the dust scattering correction for XMM--Newton and NuSTAR.}
    \label{tab:dustScatterning}
    \begin{tabular}{lccccc}
        \toprule
        \toprule
        Instrument & Region & $\Gamma_{\rm before}$ & $\Gamma_{\rm after}$ & $\rm SB_{\rm before}$ & $\rm SB_{\rm after}$ \\
         \midrule
         \multirow{3}{*}{XMM--Newton} & $r<15''$ & $1.44\pm0.11$ & $1.93\pm0.11$ & $7.4\pm0.4$ & $9.6\pm0.4$ \\
         & $15''<r<27.5''$ & $1.95\pm0.11$ & $1.95\pm0.11$ & $2.86\pm0.13$ & $2.86\pm0.13$ \\
         & $27.5''<r<40''$ & $2.09\pm0.14$ & $2.05\pm0.15$ & $1.62\pm0.09$ & $1.21\pm0.07$ \\
         \midrule
         \multirow{3}{*}{NuSTAR} & $r<24''$ & $2.00\pm0.10$ & $2.16\pm0.10$ & $4.7\pm0.3$ & $5.0\pm0.3$ \\
         & $24''<r<38''$ & $2.28\pm0.13$ & $2.35\pm0.15$ & $2.32\pm0.16$ & $2.40\pm0.17$ \\
         & $38''<r<50''$ & $2.27\pm0.14$ & $2.29\pm0.17$ & $1.56\pm0.14$ & $1.55\pm0.14$ \\
         \bottomrule
    \end{tabular}
    \tablecomments{The notation ``SB" stands for surface brightness, with units $10^{-12}\ \text{erg}\ \text{s}^{-1}\ \text{cm}^{-2}\ \text{arcmin}^{-2}$. All uncertainties are reported at the 90\% confidence level.}
\end{table*}

\subsubsection{X-Ray SED Data Binning} \label{subsubsec:cflux}
To minimize bias in the multiwavelength SED model, it is important to rebin the X-ray data to match the number of flux points per energy decade that is measured in the TeV regime. The VHE data we consider spans approximately two decades in energy from $\sim200$\,GeV to $\sim 20$\,TeV, while the X-ray band spans over one decade in energy. Therefore, we obtained the X-ray flux in roughly half of the number of TeV points using seven logarithmically spaced bins. To derive the value of the flux in each bin, we added to the joint fit model the \texttt{Xspec} convolution component \texttt{cflux}, which measures the integrated flux between two energy values. We adapted the minimum and maximum energy values that match the edges of the new bins and derived the X-ray flux. Since \texttt{cflux} computes the integral of the best-fit model, we must perform a correction to each bin flux value accordingly. For sufficiently small bins (i.e., for $\Delta E_i / E_i \ll 1$), the flux integral can be approximated to
\begin{equation}
    \begin{split}
    \texttt{cflux}_i & = \int_{E_i - \frac{\Delta E_i}{2}}^{E_i + \frac{\Delta E_i}{2}} \tilde{N}(E) E dE \approx \tilde{N}(E_i) \times E_i \times \Delta E_i \\
    & \approx N_i(E_i) \times E_i \times \Delta E_i,
    \end{split}
\end{equation}
where $\tilde N (E_i)$ and $N_i$ are the photon flux in units of ph cm$^{-2}$ s$^{-1}$\,keV$^{-1}$ for the best-fit model and the $i$-th photon flux, and we assume $\tilde N (E_i) \simeq N_i$. From this, the $i$-th SED value is corrected using the following prescription.
\begin{equation}
    \text{SED}_i = \texttt{cflux}_i \times \frac{E_i}{\Delta E_i}.
\end{equation}
The rebinned X-ray flux values are calculated for energies 2--30\,keV and are incorporated in multiwavelength SED modeling, see Sect.~\ref{sec:sed}.

\section{SED Modeling} \label{sec:sed}
We present SED modeling of PWN~G0.9+0.1, testing two different approaches: a dynamical one-zone model and a multi-zone model, and compare their results. The morphology of PWN~G0.9+0.1 in the radio indicates it has not yet interacted with its parent SNR, making it an ideal source to evaluate the different approaches commonly employed in current models. We consider the X-ray data from Sect.~\ref{subsubsec:cflux} together with the most recently available multiwavelength data: the radio data of \cite{Dubner_2008}, the VHE data observed by HESS \citep{HESS_2005}, and the VHE data observed by VERITAS \citep{VERITAS_2021}. The right panel of Fig.~\ref{fig:multiwaveImage_withSNR} displays the MeerKAT (1.28\,GHz), XMM--Newton (2--10\,keV) and NuSTAR (10--30\,keV) images of PWN~G0.9+0.1, together with the extension of the source as observed by HESS. We additionally include the Fermi--LAT upper limits of \cite{Acharyya_2025}, rebinning the data for energies above 1\,GeV.

\subsection{Dynamical One-zone Model} \label{subsec:model_dyn}
The dynamical one-zone model we adopt in this work was first presented in \cite{Gelfand_2009} and has been applied to several PWNe in different stages of evolution \citep[e.g.,][]{Burgess_2022, Woo_2023, Abdelmaguid_2023, Pope_2024, Alford_2025}. It is a time-dependent evolutionary model of a spherical PWN inside a parent SNR that expands within a uniform ISM, where the SNR evolves from freely expanding to the Sedov-Taylor phase, after which radiative cooling starts to take over the SNR evolution. The PWN magnetic field is assumed to be homogeneously distributed and isotropic inside the volume. The system is evolved from birth to the estimated true age of the system, taking into account physical parameters of the pulsar (e.g, characteristic age, spindown power), the supernova explosion (e.g., ejecta mass and energy of the explosion), the surrounding medium (e.g., ISM density), and the observed sizes of the PWN and SNR. Both the dynamical and radiative properties of the system are derived at each stage of the evolution. At each time step $t_{\rm step}$\footnote{$t_{\rm step}=N$ means the code saves the properties of the PWN every $N$-th iteration divided by time.}, we obtain the particle and photon SEDs as well as the dynamical parameters of the PWN-SNR system, such as the PWN magnetic field or expansion velocity. 
 
The evolution of the relativistic particles generating PWN emission is ruled by the balance between the continuous injection of fresh particles from the pulsar at the TS boundary and the adiabatic and radiative losses of the population. The first are due to the expansion of the PWN, the second are due to synchrotron and ICS radiative processes. The injected power by the pulsar is defined by the spindown power, which can be parameterized as: 
\begin{equation} \label{eq:edot}
    \dot{E}(t) = \dot{E_0} \Bigg( 1+ \frac{t_{\rm age}}{\tau_{\rm sd}} \Bigg)^{-\frac{n+1}{n-1}},
\end{equation}
where $\dot{E_0}$ is the spindown power of the pulsar at birth, $t_{\rm age}$ is the age of the pulsar, $\tau_{\rm sd}$ is the spindown timescale, and $n$ is the pulsar braking index. $\tau_{\rm sd}$ is connected to the true age through the relation:
\begin{equation} \label{eq:tau_sd}
    t_{\rm age} = \frac{2\tau_c}{n-1} - \tau_{\rm sd},
\end{equation}
where $\tau_c$ is the characteristic age of the pulsar defined by the pulsar period $P$ and its derivative $\dot{P}$, $\tau_c=P/2\dot{P}$. The model assumes all of the spindown power is injected into the PWN as either magnetic field energy ($\dot{E}_{\rm inj,B}=\eta_B\dot{E}$) or particle energy ($\dot{E}_{\rm inj,p}=(1-\eta_B)\dot{E}$). The injection spectrum is modeled as a power law between a minimum energy $E_{\rm min}$ and a maximum energy $E_{\rm max}$. 

The pressure balance between the PWN and the SNR generates a force on the thin shell of swept-up material by the freely-expanding PWN. From momentum conservation, the model derives the radius and velocity of the PWN at each time step. The dynamical evolution is primarily influenced by the characteristics of the supernova (SN) explosion (ejecta mass, $M_{\rm ej}$, and SN energy, $E_{\rm SN}$) and of the surrounding ISM (density of the ISM, $n_{\rm ISM}$). Both the radius of the PWN and of the SNR are estimated and compared to the observed sizes in radio, motivated by the minimal affects of synchrotron losses on the radio sizes relative to the X-ray sizes. During the free expansion phase, the size of the PWN changes in time as it keeps expanding, and, as a consequence, the magnetic field tends to decrease with age, affecting the synchrotron peak of the SED. By performing a Markov Chain Monte Carlo fitting, we find the best set of parameters that reproduces the present-day SED of PWN~G0.9+0.1 as well as the observed sizes of the PWN and SNR in the radio band. We estimated the angular diameter of the PWN to be $d_{\rm PWN}\sim 2.4'$ and the SNR shell to be $d_{\rm SNR} \sim 8'$, measured from MeerKAT at 1.28\,GHz \citep{MeerKAT_2022}. We fixed the source distance $d=8.5$\,kpc, similar to that of the GC. This value was adopted in previous works \citep[e.g.,][]{HESS_2005, Fang_Zhang_2010}, and is comparable to the distance estimated using the latest electron density model and reported in the Australia Telescope National Facility pulsar catalog\footnote{\url{https://www.atnf.csiro.au/research/pulsar/psrcat/}} \citep{Manchester_2005}. This corresponds to physical sizes $d_{\rm PWN}\sim 6$\,pc and $d_{\rm SNR} \sim 20$\,pc for the PWN and SNR, respectively. These values will later be compared to the best-fit sizes obtained from the modeling. We additionally fixed the braking index of the pulsar $n=3$, a typical value adopted for pulsar radiation and corresponding to perfect dipole emission. The minimum energy $E_{\rm min}$ was left free to vary during the fitting procedure, similarly to what has been done in \cite{Abdelmaguid_2023, Woo_2023, Alford_2025}. 

\begin{figure*}[ht]
    \centering
    \begin{subfigure}[b]{0.48\textwidth}
         \centering
         \includegraphics[width=\textwidth]{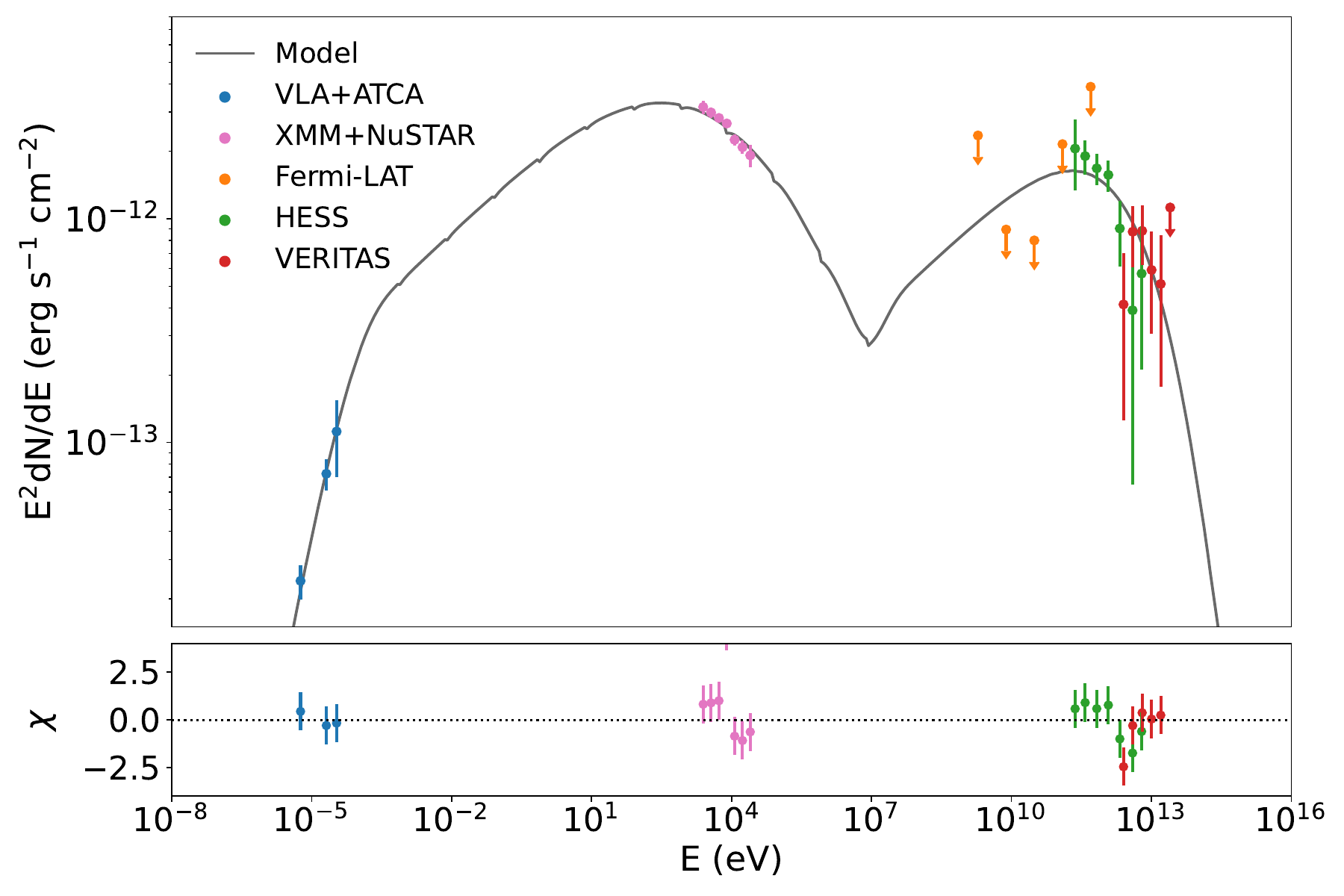}
         \caption{}
         \label{subfig:oneZone_res_SED}
     \end{subfigure}
     \begin{subfigure}[b]{0.45\textwidth}
         \centering
         \includegraphics[width=\textwidth]{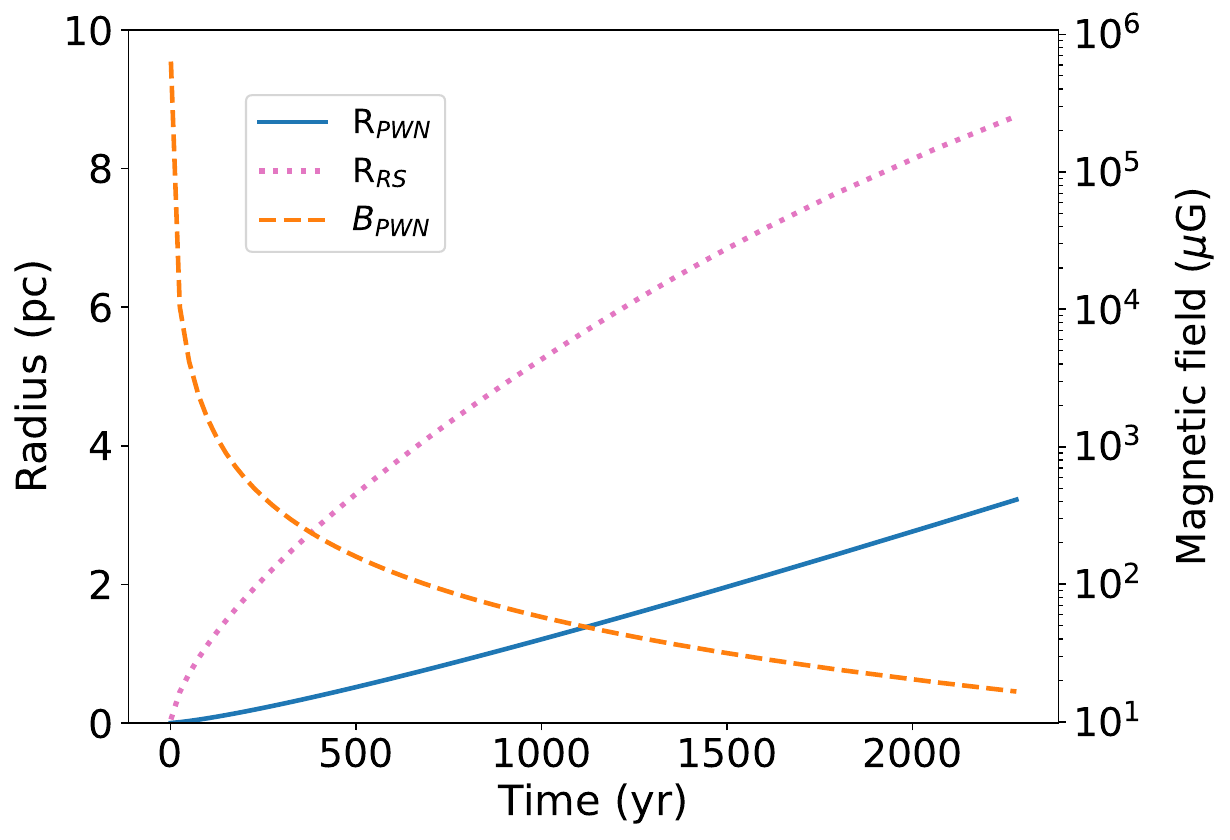}
         \caption{}
         \label{subfig:oneZone_res_dyninfo}
     \end{subfigure}
     \begin{subfigure}[b]{0.42\textwidth}
         \centering
         \includegraphics[width=\textwidth]{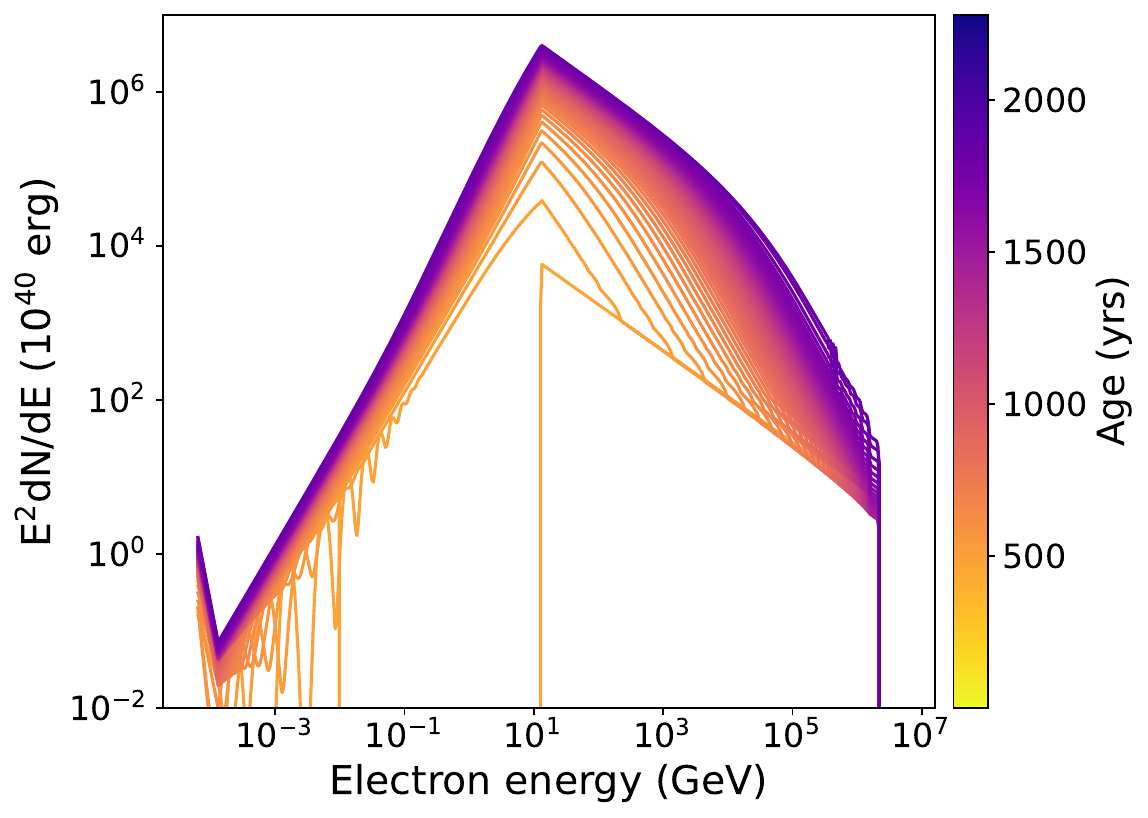}
         \caption{}
         \label{subfig:oneZone_res_particle}
     \end{subfigure}
     \hspace{10pt}
     \begin{subfigure}[b]{0.42\textwidth}
         \centering
         \includegraphics[width=\textwidth]{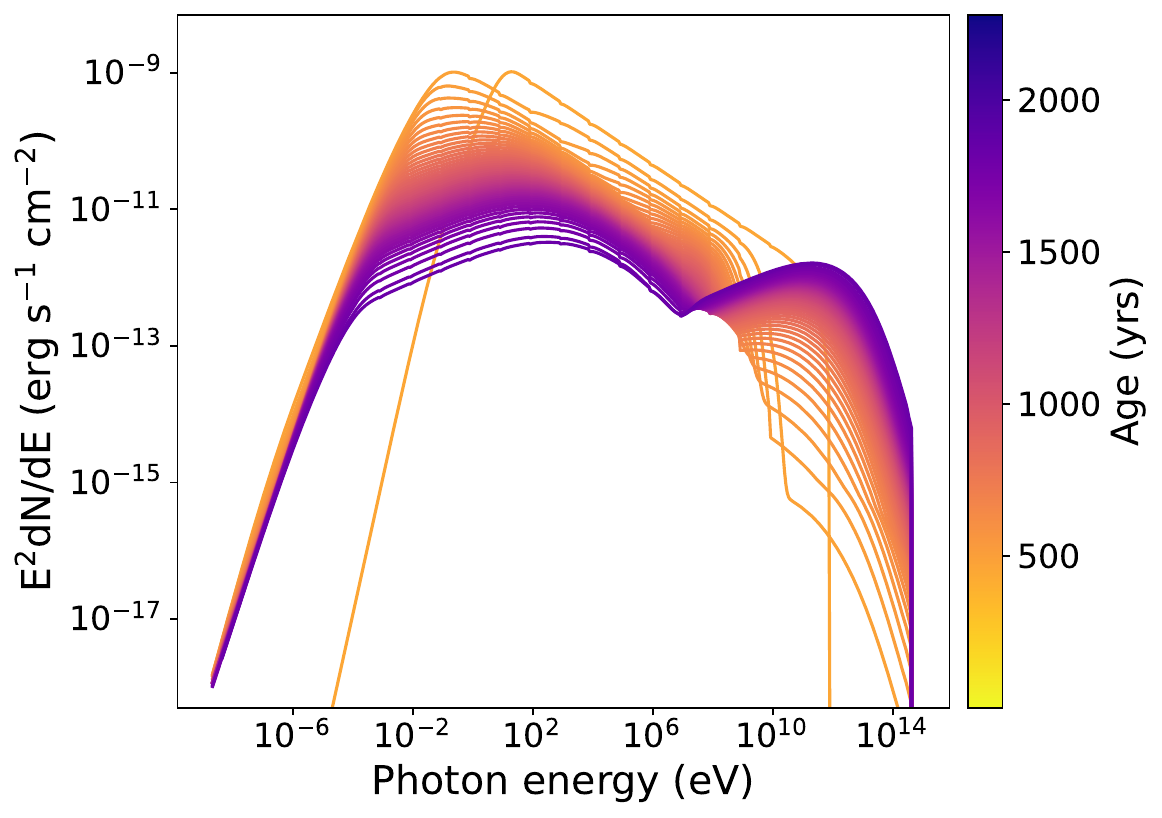}
         \caption{}
         \label{subfig:oneZone_res_photons}
     \end{subfigure}    
    \caption{Dynamical one-zone model results, obtained using a $t_{step}=50$. The model parameters are in the second column of Tab.~\ref{tab:bestfit_bothModels}. (a) Present-day SED of PWN~G0.9+0.1 fitted to the radio, X-ray, GeV, and TeV data (upper panel) along with the model residuals (bottom panel). (b) Time evolution of the PWN radius ($R_{\rm PWN}$, solid curve) and magnetic field ($B_{\rm PWN}$, dashed curve), compared to that of the RS radius of the SNR ($R_{\rm RS}$, dotted curve). The PWN has not interacted with the RS yet. (c) Time evolution of the particle spectrum. (d) Time evolution of the radiation spectrum.}
    \label{fig:oneZone_results}
\end{figure*}

The dynamical one-zone model results of the multiwavelength observations for PWN~G0.9+0.1 are shown in Fig.~\ref{fig:oneZone_results}. The top left panel displays the present-day SED model and data (Fig.~\ref{subfig:oneZone_res_SED}), the top right panel shows the evolution of the dynamical properties for the PWN (Fig.~\ref{subfig:oneZone_res_dyninfo}), and the evolution of the particle spectrum (Fig.~\ref{subfig:oneZone_res_particle}) and photon spectrum (Fig.~\ref{subfig:oneZone_res_particle}) are shown in the bottom panels. The best-fit parameters are reported in the second column of Tab.~\ref{tab:bestfit_bothModels}. The particle injection spectrum of PWN~G0.9+0.1 is well described by a single power law with a spectral index of $p \sim2.6$, a minimum energy of $E_{\rm min} \sim12.8$\,GeV, and a maximum particle energy of $E_{\rm max} \sim2.2$\,PeV. The high value for the maximum particle energy is required to match the X-ray spectrum. We explored a broken power law distribution for the particle injection spectrum, resulting in a break energy $E_{\rm break}\sim20$\,GeV, very close to the minimum energy $E_{\rm min}\sim10$\,GeV, which is essentially a single power law shape. The SN leading to PWN~G0.9+0.1 is predicted to be energetic ($E_{\rm SN} \sim 2.9 \times 10^{51}\ \text{erg s}^{-1}$, $M_{\rm ej} \sim 14\ M_\odot$) and the age of the system to be young $t_{\rm age} \sim 2.2$\,kyr. The predicted age is compatible with the lower limit measured from radio observations, $t \sim1.1$\,kyr \citep{Dubner_2008} and the estimate from X-ray observations, $t\sim 2.7$\,kyr \citep{Sidoli_2000}. From Fig.~\ref{subfig:oneZone_res_dyninfo}, it is evident that the PWN has not interacted yet with the RS of the SNR. The predicted sizes of both the PWN ($d_{\rm PWN} \sim 6.4$\,pc) and SNR ($d_{\rm SNR} \sim 21.6$\,pc) are comparable to the radio extension observed by MeerKAT ($d_{\rm PWN} \sim 6$\,pc, $d_{\rm SNR} \sim 20$\,pc). The magnetic fraction of the pulsar wind and the average PWN magnetic field are  $\eta_B \sim 0.023$ and $B_{\rm PWN} \sim 16.6\ \mu$G, respectively. The magnetic fraction and the magnetic field strength are in the typical range observed for young PWNe \citep[e.g.,][]{Tanaka_Takahara_2011, Torres_2014, Hattori_2020}. For the ICS, two photon fields are required: the CMB ($T_{\rm CMB} = 2.7$\,K, $u_{\rm CMB}=0.26$\,eV\,cm$^{3}$) and an infrared photon field ($T_{\rm IR} \sim 54$\,K and $u_{\rm IR} \sim 5.6$\,eV\,cm$^{3}$).

\begin{table*}[ht]
    \centering
    \caption{Best fit parameters for the one-zone and multi-zone models.}
    \label{tab:bestfit_bothModels}
    \begin{tabular}{lcc}
    \toprule
    \toprule
    Parameter & One-zone Model & Multi-zone Model \\
    \midrule
    \multicolumn{3}{c}{SNR parameters} \\
    \midrule
    Supernova energy $E_{\rm SN}$ ($10^{51}$ erg) & 2.9 & ... \\
    Ejecta mass $M_{\rm ej}$ ($M_{\odot}$) & 14.0 & ... \\
    ISM density $n_{\rm ISM}$ (cm$^{-3}$) & 0.23 & ... \\
    Radius $r_{\rm SNR}$(pc) & 10.8 & ... \\
    \midrule
    \multicolumn{3}{c}{Pulsar parameters} \\
    \midrule
    Age (kyr) & 2.2 & 2.2\footnote{\label{1}Held fixed to the observational value.} \\
    Spindown timescale $\tau_{\rm sd}$ (kyr) & 3.1 & 3.1\footref{1}\\
    Initial spin down power $\dot{E}_0$ ($10^{38}$ erg s$^{-1}$) & 1.3 & 1.3\footref{1}\\
    \midrule
    \multicolumn{3}{c}{PWN parameters}\\
    \midrule
    Spectral index $p$ & 2.60 & 2.61 \\
    Minimum energy $E_{\rm min}$ (GeV) & 12.8 & 13.3 \\
    Maximum energy $E_{\rm max}$ (PeV) & 2.2 & 2.0 \\
    Magnetic fraction $\eta_B$ & 0.023 & 0.009 \\
    Radius $r_{\rm PWN}$ (pc) & 3.2 & 2.1\footref{1} \\
    Velocity $V_{\rm PWN}$ & 0.005c & ... \\
    Bulk flow velocity at TS $V_0$ & ... & 0.04c \\
    Bulk flow velocity index $\alpha_V$ & ... & -0.86 \\
    Average magnetic field ($\mu$G) & 16.6 & 20.5 \\
    Magnetic field at TS $B_0$ ($\mu$G) & ... & 32 \\
    Magnetic field index $\alpha_B$ & ... & -0.14 \\
    ICS field temperature $T_{\rm IR}$ (K) &  54 & 54 \\
    ICS field energy density $u_{\rm IR}$ (eV cm$^{-3}$) & 5.6 & 2.2 \\
    \bottomrule
    \end{tabular}
    \tablecomments{The one-zone model aims at reproducing the radio size, while the multi-zone model fixes the size to that observed in the X-ray band.}
\end{table*}

\subsection{Multi-zone Model} \label{subsec:model_multizone}
The multi-zone model is a recent development described in \cite{Kim_2020} and applied to PWNe such as the ``Kookaburra" and the ``Rabbit" \citep{Park_2023a, Park_2023b}. The model follows a different approach than that of \cite{Gelfand_2009}. It does not perform the time evolution of the PWN-SNR system and describes the present-day SED of the PWN by assuming a radial dependence of the magnetic field and the bulk flow velocity of the PWN to reproduce the observed X-ray photon index and the X-ray surface brightness radial profiles. \cite{Gelfand_2009}, on the other hand, assumes a constant magnetic field and expansion velocity throughout the PWN. As in \cite{Gelfand_2009}, the spindown power of the pulsar, defined by Eq.~\ref{eq:edot}, is assumed to be injected into the PWN as magnetic field energy ($\dot{E}_{\rm inj,B}=\eta_B\dot{E}$) and particle energy ($\dot{E}_{\rm inj,p}=\eta_p\dot{E}$). Pulsar gamma-ray radiation ($\dot{E}_{\rm inj,\gamma}=\eta_{\gamma}\dot{E}$) can be considered, but in this case, we assume $\eta_{\gamma}=0$, as in Sect.~\ref{subsec:model_dyn}, since the central pulsar is not known to emit gamma rays. Particles are injected at the TS with a power law or broken power law spectrum with Lorentz factors (energy) between $\gamma_{\rm min}$ ($E_{\rm min}$) and $\gamma_{\rm max}$ ($E_{\rm max}$). The particles flow into a spherically symmetric PWN, losing energy via advection and diffusion, and via synchrotron and ICS radiation until the age of the system is reached. The diffusion coefficient of the particles is modeled as:
\begin{equation}
    D = D_0 \Bigg( \frac{100\ \mu G}{B} \Bigg) \Bigg( \frac{\gamma_e}{10^9} \Bigg)^{\alpha_D},
\end{equation}
where $\alpha_D=1/3$ assuming Kolmogorov diffusion, $B$ is the magnetic field of the PWN and $\gamma_e$ is the Lorentz factor of the electrons. The magnetic field and velocity radial profiles of the PWN are characterized as a power law. The magnetic field is prescribed as
\begin{equation}
    \label{eq:bfield}
    B(r)=
    \begin{cases}
        B_0 \left( \frac{r}{R_{\rm TS}} \right)^{\alpha_B}, & \textrm{for } R_{\rm TS} \le r \le R_{\rm PWN} \\
        B_{\rm ext}, & \textrm{for } R_{\rm PWN} < r \le R_{\rm max}, \\
    \end{cases}
\end{equation}
where $B_0$ is the magnetic field at the TS, $R_{\rm TS}$ is the TS radius, $\alpha_B$ is the magnetic field index, $B_{\rm ext}$ is the magnetic field outside the PWN, and $R_{\rm max}$ is the maximum radius. Similarly, for the bulk flow velocity:
\begin{equation}
    \label{eq:vflow}
    V_{\rm flow}(r)=
    \begin{cases}
        V_0 \left( \frac{r}{R_{\rm TS}} \right )^{\alpha_V}, & \textrm{for } R_{\rm TS} \le r \le R_{\rm PWN}, \\
        V_{\rm ext}, & \textrm{for } R_{\rm PWN} < r \le R_{\rm max}, \\
    \end{cases}
\end{equation}
where $V_0$ is the bulk flow speed at the TS, $\alpha_V$ is the bulk flow velocity index, and $V_{\rm ext}$ is the velocity outside the PWN. The power law indices satisfy $\alpha_B+\alpha_V=-1$, which is valid for a spherical geometry of the flow if magnetic flux is conserved \citep{Reynolds_2009}. The described model reproduces the observed SED, X-ray photon index, and the X-ray surface brightness radial profiles measured by XMM--Newton and NuSTAR. We fixed the distance to $d=8.5$\,kpc and age to $t=2.2$\,kyr. The PWN radius was fixed to $r=50''=2.1$\,pc, based on the observed X-ray extent for a distance of 8.5\,kpc. The TS was assumed to be $R_{\rm TS}=0.1$\,pc, as observed in most of PWNe \citep{Kargaltsev_2008}. The maximum radius, instead, was set to $R_{\rm max}=0.1^\circ$ ($\sim15$\,pc for a distance of 8.5\,kpc) to approximately match the VERITAS positional uncertainty. Similarly to the one-zone model, the minimum energy was left free to vary during the fitting procedure, similarly to \cite{Park_2023a, Park_2023b, Kim_2024}.

\begin{figure*}[ht]
    \centering
    \begin{subfigure}[b]{0.48\textwidth}
         \centering
         \includegraphics[width=\textwidth]{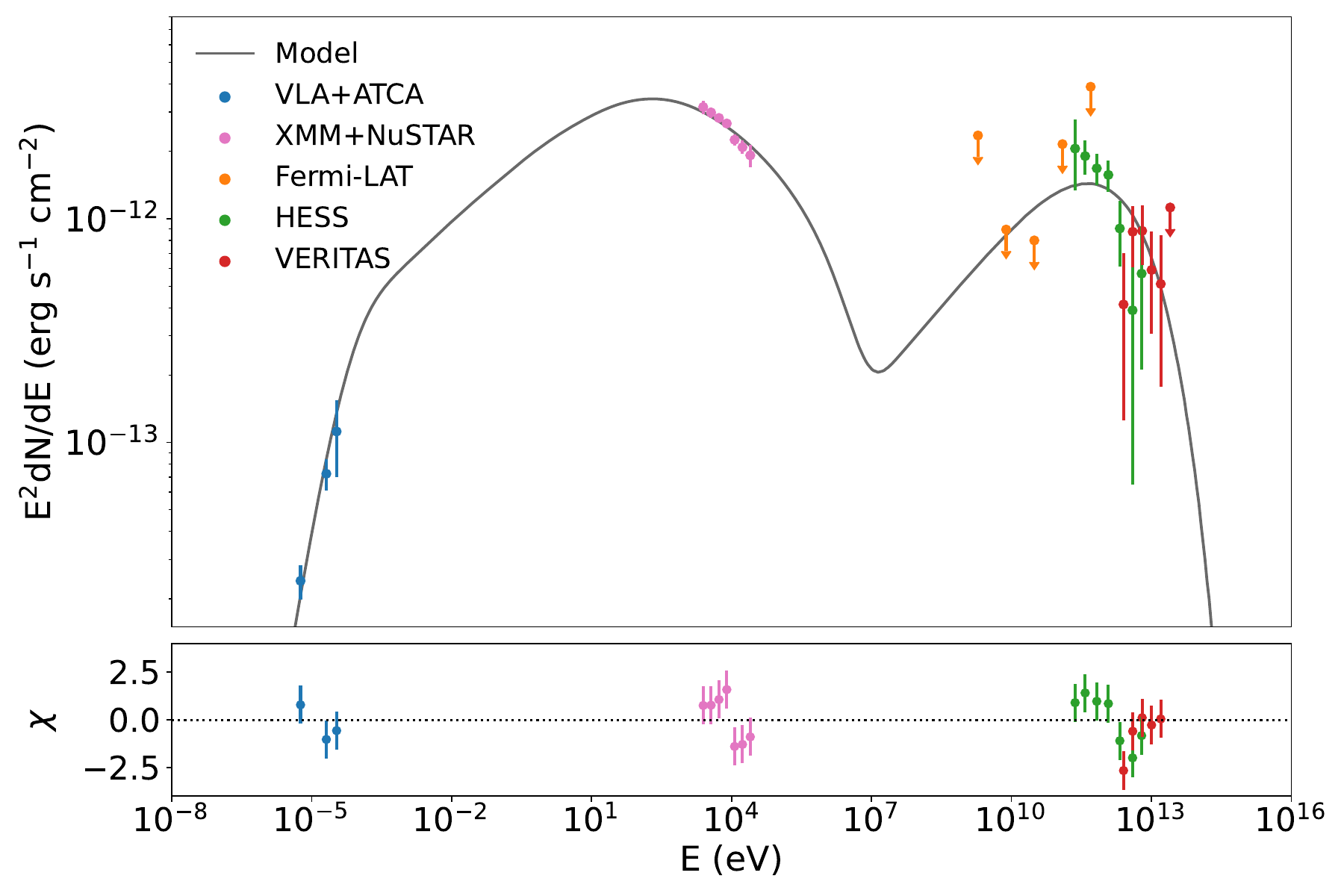}
         \caption{}
         \label{subfig:multiZone_res_SED}
     \end{subfigure}
     \\
     \begin{subfigure}[b]{0.4\textwidth}
         \centering
         \includegraphics[width=\textwidth]{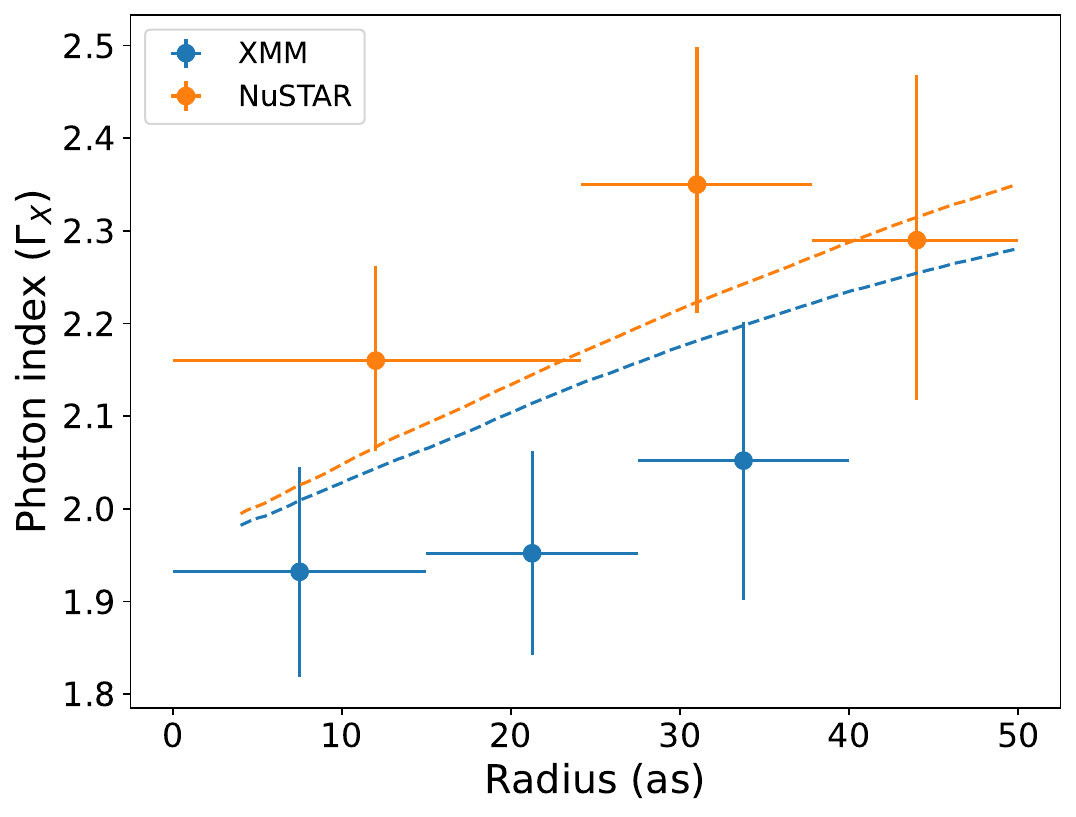}
         \caption{}
         \label{subfig:multiZone_res_gamma}
     \end{subfigure}
     \hspace{10pt}
     \begin{subfigure}[b]{0.43\textwidth}
         \centering
         \includegraphics[width=\textwidth]{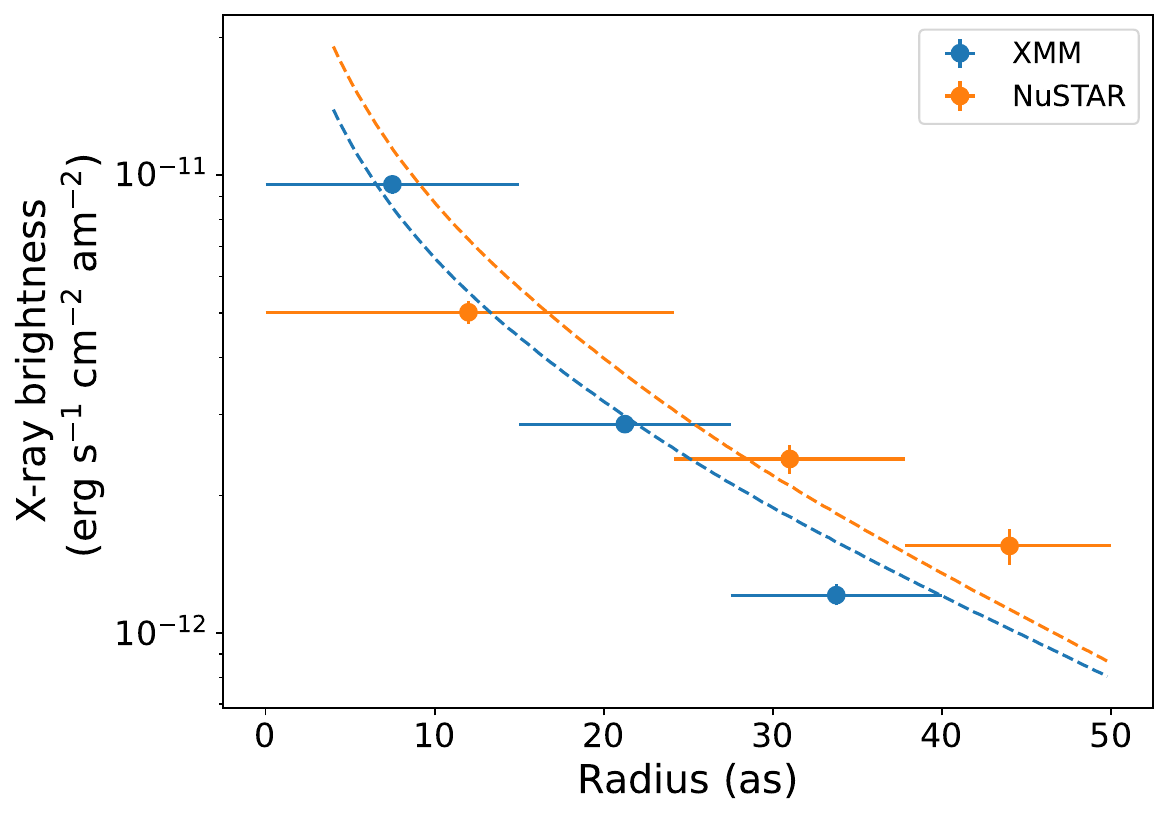}
         \caption{}
         \label{subfig:multiZone_res_SB}
    \end{subfigure}    
    \caption{Multi-zone model results. The model parameters are in the final column of Tab.~\ref{tab:bestfit_bothModels}. (a) Present-day SED of PWN~G0.9+0.1, including the radio, X-ray, GeV, and TeV data (upper panel) along with the model residuals (bottom panel). (b) Radial profile data and model of the X-ray photon index. (c) Radial profile data and model of the X-ray surface brightness.}
    \label{fig:multiZone_results}
\end{figure*}

The results of the multi-zone model are shown in Fig.~\ref{fig:multiZone_results}. The present-day SED model and data are shown in Fig.~\ref{subfig:multiZone_res_SED}, the fits of the X-ray photon index radial profiles are in Fig.~\ref{subfig:multiZone_res_gamma}, and the fits of the X-ray surface brightness radial profiles are in Fig.~\ref{subfig:multiZone_res_SB}. The best fit parameters are reported in the last column of Tab.~\ref{tab:bestfit_bothModels}. The SED can be well described as a single power law particle injection spectrum with an index $p\sim2.61$ and a minimum and maximum energy of $E_{\rm min}\sim13.3$\,GeV and $E_{\rm max} \sim2.0$\,PeV. We found similar results for a broken power law particle injection spectrum as in Sect.~\ref{subsec:model_dyn}, $E_{\rm min}\sim10$\,GeV and $E_{\rm break}\sim 30$\,GeV, comparable to a single power law. The magnetic fraction is $\eta_B \sim 10^{-2}$. The average magnetic field value, obtained from averaging $B(r)$ between the TS radius and the PWN size, is $B_{\rm PWN}\sim20.5\ \mu$G. The average bulk flow velocity is $\bar{V}\sim0.004c$. An additional photon field to the CMB is required for the ICS emission to characterize the gamma-ray data sufficiently, resulting in an infrared field with $T_{\rm IR}=54$\,K. To reproduce the X-ray radial profiles, a diffusion coefficient $D_0=1 \times 10^{27}$\,cm$^2$ s$^{-1}$ and a velocity index $\alpha_V=-0.86$ are required which yields $\alpha_B=-0.14$. The XMM--Newton and NuSTAR surface brightness profiles in Fig.~\ref{subfig:multiZone_res_SB} are well fitted by the model, though NuSTAR shows a slight discrepancy in the last point corresponding to $r \sim 45''$, possibly explained by the higher contamination from background, stray light that affects NuSTAR more than XMM--Newton. The model reasonably fits the photon index profiles, with deviations that could be explained by the assumption of spherical symmetry for the PWN, which is observed to have a more complex torus-jet morphology in the Chandra data. Deeper Chandra observations would enable a spectral study of the torus and jet regions. Alternatively, the X-ray photon index profiles could be better characterized using a harder particle injection spectrum with a lower $E_{\rm max}$, but at the cost of a less accurate characterization of the observed SED, particularly in the hard X-ray band.

\section{The Very Faint X-ray Transient XMMU J174716.1--281048} \label{sec:xmmu}
The new NuSTAR observation of PWN~G0.91+0.1 has detected a second X-ray source in the FoV. The position is consistent with the VFXT XMMU~J174716.1--281048, serendipitously discovered in 2003 by XMM--Newton \citep{Sidoli_2003_ATel}. \cite{delSanto_2007a} classified the source as a VFXT that hosts a neutron star. The source distance is $\sim8$\,kpc \citep{delSanto_2007b}. The 2003 XMM--Newton spectrum was best fitted by an absorbed power law with column density $N_H=(8.9 \pm 0.5) \times 10^{22}\ \text{cm}^{-2}$, a photon index $\Gamma = 2.1 \pm 0.1$, and an unabsorbed 2--10\,keV flux $F_{2-10\text{keV}}=(6.8 \pm 0.4)\times 10^{-12}\ \text{erg}\ \text{cm}^{-2}\ \text{s}^{-1}$ \citep{delSanto_2007a}. The flux corresponds to a 2--10\,keV luminosity of $L_X \sim 5 \times 10^{34}$ erg s$^{-1}$. A type I X-ray burst was observed from the VFXT in 2010 \citep{Degenaar_2011}, with a comparable luminosity to the 2003 event, before entering a quiescent state around 2015 with a 2--10\,keV luminosity $L_X \lesssim 4.6 \times 10^{33}$\,erg s$^{-1}$ \citep{delSanto_2015_ATel}. The NuSTAR result we report here is the first detection of the VFXT above 10\,keV.

We extracted the NuSTAR spectrum from a $30''$ circular region and fitted it in the 3--30\,keV band with an absorbed power law. The best fit parameters are $N_H = (1.0 \pm 0.5) \times 10^{23}\ \text{cm}^{-2}$ and $\Gamma = 1.83 \pm 0.15$, for a fit value $\chi^2=131.9$ for 143 dof ($\chi^2/dof=0.92$). The corresponding unabsorbed flux is $F_{3-30\text{keV}}=(8.1 \pm 0.5)\times 10^{-12}\ \text{erg}\ \text{cm}^{-2}\ \text{s}^{-1}$. We also reanalyzed the 2003 XMM--Newton dataset in the range 0.5--10\,keV, and obtained spectral results consistent with those of \cite{delSanto_2007a}. The 3--10\,keV flux measured with XMM--Newton is $F_{\rm XMM}(3-10 \ \text{keV}) = (4.49 \pm 0.12)\times 10^{-12}\ \text{erg}\ \text{cm}^{-2}\ \text{s}^{-1}$. While the NuSTAR flux is slightly lower, $F_{\rm NuSTAR}(3-10 \ \text{keV})=3.8^{+0.6}_{-0.5}\times 10^{-12}\ \text{erg}\ \text{cm}^{-2}\ \text{s}^{-1}$, the two values are still compatible within the statistical uncertainties. When fixing the column density in the NuSTAR model to the XMM--Newton derived value, the resulting 3--10\,keV flux is more consistent with the XMM--Newton measurement, $F_{\rm NuSTAR}(3-10 \ \text{keV}) = (4.2 \pm 0.3) \times 10^{-12}\ \text{erg}\ \text{cm}^{-2}\ \text{s}^{-1}$. The corresponding 3--10\,keV luminosities at 8\,kpc are $L_X = (3.4 \pm 0.1) \times 10^{34}$\,erg s$^{-1}$ and $L_X = (3.2 \pm 0.2) \times 10^{34}$\,erg s$^{-1}$ for XMM--Newton and NuSTAR, respectively. It is interesting to note that the luminosity measured by NuSTAR in 2021 is compatible with the value obtained by XMM--Newton in 2003, when the source was reported to be quiescent in 2015. It appears XMMU~J174716.1--281048 is exiting its quiescent state, and is likely back to an outbursting state as at the time of its discovery in 2003.

\begin{figure}[hbt]
    \centering
    \includegraphics[width=1.0\linewidth]{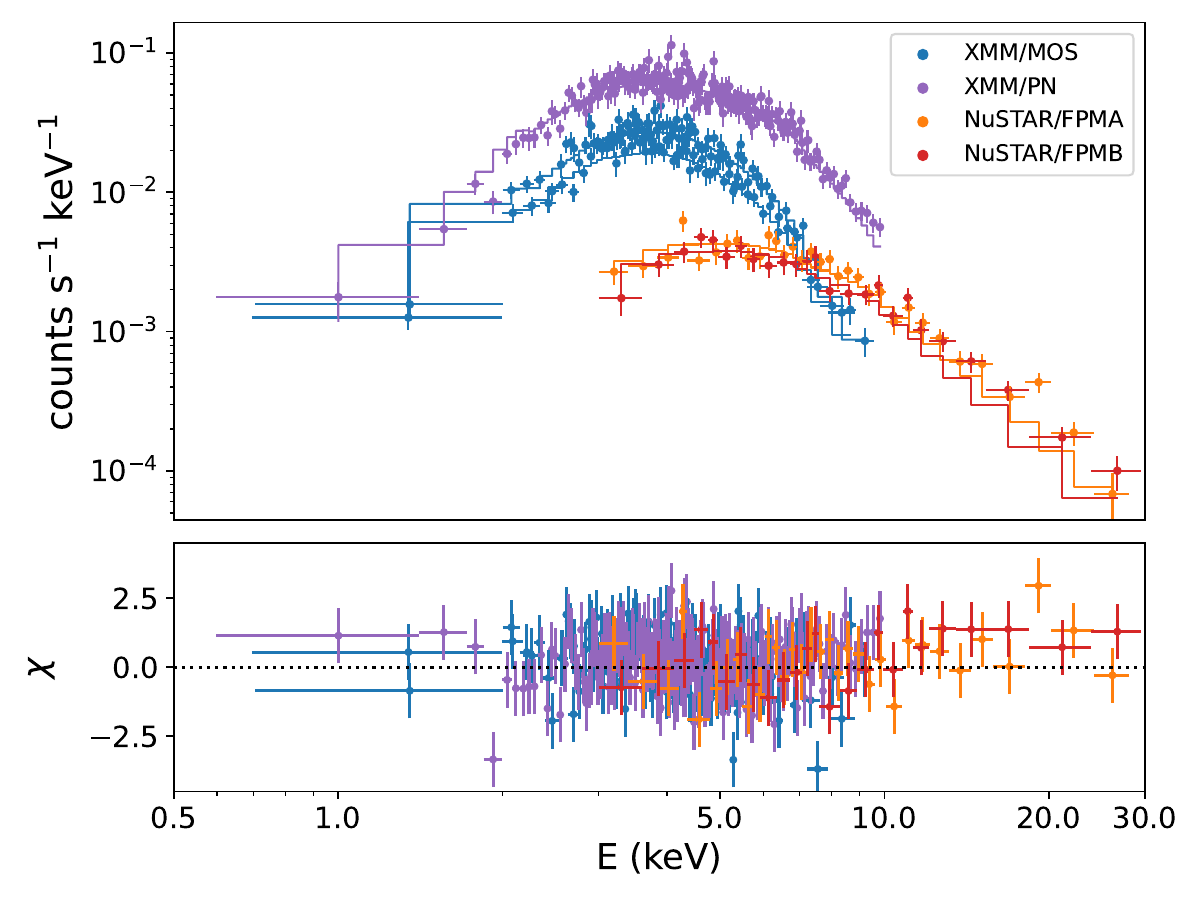}
    \caption{Joint fit of both XMM--Newton and NuSTAR spectral data for the VFXT XMMU~J174716.1--281048 in the energy range 0.5--30\,keV using an absorbed power law model (upper panel), along with the model residuals (bottom panel) expressed as (data - model)/error.}
    \label{fig:jointFit_xmmu}
\end{figure}

A joint fit of the XMM--Newton (0.5--10\,keV) and NuSTAR (3--30\,keV) spectra provided a statistically acceptable result ($\chi^2/dof=0.99$) with best fit parameters $N_H = (12.4 \pm 0.4) \times 10^{22}\ \text{cm}^{-2}$, $\Gamma = 2.07 \pm 0.07$ and a 3--10\,keV unabsorbed flux $F_{3-10} = 4.45^{+0.12}_{-0.10}\times 10^{-12}\ \text{erg}\ \text{cm}^{-2}\ \text{s}^{-1}$. The best-fit model is shown in Fig.~\ref{fig:jointFit_xmmu} for the 0.5--30\,keV range. The corresponding X-ray luminosity at 8\,kpc is $L_X=3.41^{+0.09}_{-0.08} \times 10^{34}$\,erg s$^{-1}$. We selected the pn spectrum as the reference because of its better signal-to-noise. The cross-calibration constants are $c_{\rm M}= 1.04 \pm 0.03$, $c_{\rm N}=0.97 \pm 0.05$, for MOS1 and 2 from XMM--Newton and NuSTAR FPMA and FPMB, respectively. We did not consider the dust scattering effect since the distance to the source is not well determined.

\section{Discussion} \label{sec:discussion}
\subsection{Comparison Between Models} \label{subsec:model_comparison}
Both models employed in this work use a similar set of assumptions: the injection history from the pulsar, described by Eq. (\ref{eq:edot}), and the spindown power is completely converted into either particle or magnetic field energy. In both models, the PWN is assumed to be a spherically symmetric bubble of relativistic electrons and positrons that undergo synchrotron and ICS radiative losses. The main difference in the two approaches is the treatment of particle propagation and the time evolution of the PWN. The dynamical one-zone model \citep{Gelfand_2009} is time-dependent and assumes the PWN properties as uniform, and reproduces the present-day SED and the observed size of both the PWN and the SNR. The multi-zone model \citep{Kim_2020} does not consider time evolution, but invokes a radial dependence of both the magnetic field and the bulk flow velocity of the PWN, and reproduces the present-day SED as well as X-ray radial profiles. In this study, we demonstrate that both models can reproduce well the observed SED and X-ray properties of PWN~G0.9+0.1 with best fit parameters that are generally agreeable and are reported in Tab.~\ref{tab:bestfit_bothModels}. Both models predict that the injected particle spectrum can be well described by a single power law with a soft spectral index of $p\sim 2.6$, a minimum energy $\sim13$\,GeV, and a magnetic fraction $\eta_B \sim (0.9 - 2) \times 10^{-2}$. More importantly, both models find a maximum particle energy $E_{\rm max}\sim2$\,PeV to accurately characterize the X-ray data, especially in the hard $E> 10$\,keV band. Overall, we can constrain the magnetic field to $B_{\rm PWN}\sim 20\ \mu$G, where the differences between the two models can be explained by the different treatment of the magnetic field.

The average value of the bulk flow velocity for the multi-zone model ($\bar{V}\sim0.004c$, see Sect.~\ref{subsec:model_multizone}) is close to the average velocity of the PWN in the one-zone model ($V_{\rm PWN} \sim 0.005c$, see Tab.~\ref{tab:bestfit_bothModels}). However, the two models define the velocity in different ways: $V_{\rm PWN}$ is the expansion velocity of the PWN boundary in the one-zone model, while it is defined as the bulk flow velocity inside the PWN in the multi-zone model. The injection spectrum of PWN~G0.9+0.1 is shown to be well fitted by a single power law rather than a broken power law, as most previous SED models assumed \citep{Tanaka_Takahara_2011, Torres_2014}. It is still not well understood why most PWNe might exhibit a broken power law spectrum while others, such as the ``Dragonfly" \citep{Woo_2023} and PWN~G54.1+0.3 \citep{Alford_2025}, indicate single power law spectra. This behavior probably indicates differences in the acceleration regions and mechanisms, which might also depend on the age or environment of the PWN.

The comprehensive gamma-ray data from the Fermi--LAT, HESS, and VERITAS are consistent among the different instruments, but the upper limits from the Fermi--LAT data are difficult to reproduce for either model. There are a few possible explanations for this. One possibility is that the Fermi--LAT data are not entirely free of Galactic background emission, either from the diffuse background or from the bright GC, though this appears to contradict the relatively stringent upper limits for $\sim 10-30$\,GeV, where the background should be less prominent and where the LAT is most sensitive and has the best angular resolution. Other possibilities include additional complexities to the particle population, such as a different particle distribution shape or perhaps an additional particle distribution. Future analysis investigating the gamma-ray emission in this region is required to better understand any background contribution. Future facilities such as the CTAO will provide the perfect opportunity to perform more detailed studies in the VHE band.

\subsection{PWN G0.9+0.1 as a PeVatron candidate} \label{subsec:pevatron}
The maximum particle energy predicted by the presented models, $E_{\rm max}\sim2$\,PeV, is compelling evidence that PWN~G0.9+0.1 could be a PeVatron accelerator. In Appendix~\ref{app:sed_plots_extra} we show how decreasing the value of $E_{\rm max}$ in the models results in a worse fit, in particular in the X-ray band. Hard X-ray photons are produced via synchrotron radiation from GeV--PeV electrons and are not affected by Klein-Nishina effects like ICS emission at TeV energies. As such, the new NuSTAR observation of PWN~G0.9+0.1 provides, for the first time, the necessary observational constraints to estimate $E_{\rm max}$ in the PeV regime. Previous models \citep[e.g.,]{Tanaka_Takahara_2011,Torres_2014} predicted maximum energies in the range $\sim400-900$\,TeV, and, more recently, \cite{Fiori_2020} constrained it to be $>600$\,TeV. G0.9+0.1 should be considered a new PeVatron PWN candidate, adding to the growing sample of sources previously detected with NuSTAR and modeled with either the one-zone or the multi-zone SED models used in this work. The candidate PeVatron PWNe and their parameters are listed in Tab.~\ref{tab:pwn_comparison}. All the sources are powered by energetic ($\dot{E}\sim10^{36-37}$\,erg s$^{-1}$) pulsars, but they display a variety of ages, distances, and morphologies. Most of the sources are characterized by relatively low magnetic fields and seem to have already interacted with the parent SNR. The modeling of ``Dragonfly" and ``Boomerang" proved that both PWNe have already been crushed by the passing of the RS, while the study of HESS~J1640--465 showed it has recently started interacting with it. The one-zone modeling of the ``Eel" found that the RS did not hit the PWN yet; however, the non-spherical morphology of the source suggests that it may be approaching or that the PSR is moving supersonically, elongating the structure of the PWN. The multi-zone model does not take into account the interaction with the SNR, however, the asymmetric morphologies of both the ``Kookaburra" and ``Rabbit" PWNe likely suggest that there might be some recent interaction with the RS. The spherical radio morphology of PWN~G0.9+0.1, as well as the one-zone modeling, proved that the source did not interact with the RS of the SNR and this is likely connected to the higher magnetic field ($B_{\rm PWN} \sim 20\ \mu$G) compared to the other sources ($B_{\rm PWN} \lesssim 10\ \mu$G). The magnetic field value is similar to that obtained for PWN~G21.5--0.9, another similar young ($t \sim 1.7$\,kyr) PWN for which the radio size is comparable to the X-ray one \citep{Hattori_2020}. However, the maximum energy of PWN~G21.5--0.9 is estimated to be $\sim200$\,TeV, possibly due to the non-standard characteristics of the injection spectrum, which displays $p_1> p_2$ \citep{Hattori_2020}. Overall, PWN~G0.9+0.1 can be considered a unique source within the sample of analyzed PeVatron PWNe.

\begin{table*}[ht]
\centering
\caption{Comparison between the candidate PeVatron PWNe observed with NuSTAR.}
\label{tab:pwn_comparison}
    \begin{tabular}{lccccccc}
        \toprule
        \toprule
        \multirow{2}{*}{PWN} & \multirow{2}{*}{G0.9+0.1} & G18.5-0.4\textsuperscript{a}  & G75.23+0.12\textsuperscript{b} & G106.65+2.96\textsuperscript{c} & G313.54+0.23\textsuperscript{d} & G313.3+0.1\textsuperscript{e} & G338.3--0.0\textsuperscript{f} \\
        &  & ``Eel" & ``Dragonfly'' & ``Boomerang" & ``Kookaburra" & ``Rabbit" & HESS J1640--465 \\
        \midrule
        PSR & J1747--2809 & J1825--1256 & J2021+3651 & J2220+6114 & J1420--6048 & J1418--6058 & J1640--4631 \\
        $\tau_c$ (kyr) & 5.3 & 14.4 & 17 & 10 & 13 & 10.4 & 3.4 \\
        $\dot{E}$ (erg s\textsuperscript{-1}) & $4.3\times 10^{37}$ & $3.6 \times 10^{36}$ & $3.4\times 10^{36}$ & $2.2 \times 10^{37}$ & $1 \times 10^{37}$ & $5\times 10^{36}$ & $4.4\times 10^{36}$ \\
        $d$\textsuperscript{g} (kpc) & 8--16 (8) & 3.5 & 0.4--12 (3.5) & 0.8--7.5 (7.5) & 5.6--7.7 (5.6) & 3.5 & 12 \\
        \midrule
        $E_{\rm max}$ (PeV) & 2.0--2.2 & 2 & 1.4 & 1.2 & 0.9 & 1 & 1.2 \\
        $B_{\rm PWN}$ ($\mu$G) & 20 & 1 & 2.7 & 2.5 & 5.1 & 12.3 & 4.1 \\
        $t_{\rm age}$ (kyr) & 2.2 & 5.7 & 16 & 2.1 & 9 & 7 & 3.1 \\
        Interaction & \multirow{2}{*}{No} & \multirow{2}{*}{Possible} & \multirow{2}{*}{Yes} & \multirow{2}{*}{Yes} & \multirow{2}{*}{Possible} & \multirow{2}{*}{Possible} & \multirow{2}{*}{Yes} \\
        with RS? &  &  &  &  &  &  &  \\
        \midrule
        \multirow{2}{*}{SED model} & One-zone & \multirow{2}{*}{One-zone} & \multirow{2}{*}{One-zone} & \multirow{2}{*}{One-zone} & \multirow{2}{*}{Multi-zone} & \multirow{2}{*}{Multi-zone} & \multirow{2}{*}{One-zone} \\
         & Multi-zone &  &  &  &  &  &  \\
        \bottomrule
    \end{tabular}
    \tablecomments{\textsuperscript{a}\cite{Burgess_2022};\textsuperscript{b}\cite{Woo_2023}; \textsuperscript{c}\cite{Pope_2024}; \textsuperscript{d}\cite{Park_2023a}; \textsuperscript{e}\cite{Park_2023b}; \textsuperscript{f}\cite{Abdelmaguid_2023}; \textsuperscript{g} Range of estimated distances, and assumed distance in the SED modeling in parentheses.}
\end{table*}

\section{Summary}
We have presented a detailed broadband X-ray study using XMM--Newton and a new NuSTAR observation of the PWN inside the composite SNR~G0.9+0.1. We have modeled the SED from radio to TeV gamma-rays using two different models to determine the physical properties of the system. The PWN is detected by NuSTAR up to 30\,keV as an extended source and exhibits signs of synchrotron burnoff effect. The broadband 2--30\,keV X-ray spectrum can be well described by a highly absorbed ($N_H \sim 2 \times 10^{23}\ \text{cm}^2$) power law with a photon index of $\Gamma=2.11 \pm 0.07$. We performed a spatially resolved analysis for both XMM--Newton and NuSTAR data and derived the radial profiles of the X-ray photon index and surface brightness, correcting them for the effect of dust extinction.

We also reported the NuSTAR serendipitous detection of the VFXT XMMU~J174716.1--281048, the first above 10\,keV. The source was detected during renewed X-ray activity after being in a quiescent X-ray state since 2015.

Previous radio and X-ray estimates of the age for PWN~G0.9+0.1 is $t\sim 1.1-3$\,kyr. Radio observations indicate the PWN has likely not interacted with the SNR yet, representing an ideal case to evaluate the performance of different SED models: the one-zone dynamical model by \cite{Gelfand_2009} and the multi-zone model by \cite{Kim_2020}. Both models can reproduce the present-day SED of PWN~G0.9+0.1 with compatible parameters. They both require a single power law injection spectrum with a spectral index of $p\sim2.6$ and a maximum energy $E_{\rm max}\sim2$\,PeV to match the X-ray spectrum, suggesting PWN~G0.9+0.1 is a PeVatron accelerator. The models estimate the age of the system $t \sim2.2$\,kyr, which is compatible with the radio and X-ray based values. The one-zone model we presented supports that the PWN has not yet interacted with the reverse shock of the parent SNR. The average magnetic field value of the PWN was determined to be $B_{\rm PWN}\sim 20\ \mu$G for both models.

The model comparison performed here can be expanded to a wider sample of PWNe of different ages and morphologies, with the aim of conducting a systematic study that informs us of the model performances when applied to sources at different evolutionary stages of a PWN's life. Additionally, next-generation X-ray facilities will provide deeper insights into the relativistic particle populations. Deeper Chandra or the future mission AXIS \citep{Reynolds_2023_axis} observations will be able to resolve small-scale structures within the torus and jet of PWNe like G0.9+0.1 and measure the spectral properties of the different components. Observations with the future NewAthena telescope \citep{Cruise_2025} will help us detect significant X-ray emission of the SNR shell observed in radio, as well as improve the spatially-resolved spectral study of the PWN and that of PSR~J1747--2809. Morphological studies will also be possible in the gamma-ray band, thanks to the improved angular resolution of the upcoming CTAO \citep{Zanin_2021}. Future hard X-ray missions will also improve the existing $E>10$\,keV observations \citep[e.g.,][]{Mori_2023, Reynolds_2023_hexp}, enabling better constraints on the maximum particle energy to identify new Galactic PeVatrons.

\begin{acknowledgments}
    G.B. acknowledges financial support from the University of Bologna through the Marco Polo Program and thanks the Columbia Astrophysics Laboratory at Columbia University for the hospitality during her research stay. Support for this work and K.M. was partially provided by NuSTAR AO-6 Large Program grant (80NSSC21K0042). G.P. acknowledges financial support from the European Research Council (ERC) under the European Union’s Horizon 2020 research and innovation program HotMilk (grant agreement No. 865637), and support from Bando per il Finanziamento della Ricerca Fondamentale 2022 dell’Istituto Nazionale di Astrofisica (INAF): GO Large program and from the Framework per l’Attrazione e il Rafforzamento delle Eccellenze (FARE) per la ricerca in Italia (R20L5S39T9). H.A. acknowledges support by the National Research Foundation of Korea (NRF) grant funded by the Korean Government (MSIT) (RS-2023-NR076359). This paper employs a list of Chandra datasets, obtained by the Chandra X-ray Observatory, contained in~\dataset[DOI: https://doi.org/10.25574/cdc.508]{https://doi.org/10.25574/cdc.508}. The MeerKAT telescope is operated by the South African Radio Astronomy Observatory, which is a facility of the National Research Foundation, an agency of the Department of Science and Innovation.
\end{acknowledgments}

\facilities{NuSTAR, XMM--Newton, Chandra.}

\software{
    SAS v22.1 \citep{Gabriel_2004},
    NuSTARDAS v2.2.0,
    HEASoft v6.35 \citep{HEASARC_2014},
    CIAO v4.17 \citep{Fruscione_2006},
    Xspec v12.15 \citep{Arnaud_1996},
    Sherpa \citep{Freeman_2001}
}

\appendix
\restartappendixnumbering
\section{XMM--Newton analysis details} \label{app:xmm_exta}
We used the same methodology as described in Sect.~\ref{subsubsec:xmm_spec_analysis} to analyze the two XMM--Newton epochs individually. For the 2000 observation, we obtained a $\chi^2/dof=0.78$ for 150 dof. The best fit parameters, including the MOS calibration constant, are: $N_H=(1.9 \pm 0.2)\times 10^{23}\ \text{cm}^{-2}$, $\Gamma=1.7\pm0.2$, and $c=0.96 \pm 0.06$. The 3--10\,keV flux is $F_{3-10}= (3.1 \pm 0.2)\times 10^{-12}\ \text{erg}^{-2}\ \text{s}^{-1}\ \text{cm}^{-2}$. Within statistical uncertainties, our results do not agree with those of \cite{Porquet_2003}, but may be explained by the much larger spectral extraction region they used, a $\sim2' \times 2'25''$ ellipse. The larger region washes out the presence of variations in the photon index and surface brightness as a function of distance from the pulsar. The analysis of the 2003 observation returned a $\chi^2/dof=0.895$ (393 dof) for a photon index of $\Gamma=1.89\pm0.13$ and a column density of $N_H=1.99^{+0.12}_{-0.13} \times 10^{23}\ \text{cm}^{-2}$. The MOS cross-calibration constant is $c=0.96 \pm 0.06$. The 3--10\,keV flux is $F_{3-10}= 3.38^{+0.16}_{-0.14}\times 10^{-12}\ \text{erg}^{-2}\ \text{s}^{-1}\ \text{cm}^{-2}$. The results of the two observations agree with each other within statistical uncertainties.

We also performed the joint XMM--Newton and NuSTAR fit, separating the two epochs. We used XMM--Newton data in the 2--10\,keV range and NuSTAR data in the 3--30\,keV range. As shown in Tab.~\ref{tab:app_jointFit}, for both epochs, we obtained $\chi^2/dof \sim 0.9$ and a photon index $\Gamma \sim 2.2$. The 3--10\,keV fluxes, $F_{3-10}= (3.4 \pm 0.2)\times 10^{-12}\ \text{erg}^{-2}\ \text{s}^{-1}\ \text{cm}^{-2}$ (2000) and $F_{3-10}= 3.6^{+0.1}_{-0.2}\times 10^{-12}\ \text{erg}^{-2}\ \text{s}^{-1}\ \text{cm}^{-2}$ (2003), are consistent. The cross-calibration constants are within 10\% fluctuations from the reference spectrum, which was set to the pn data set in both cases.

\begin{table*}[ht]
    \centering
    \caption{Summary of the results for the joint XMM--Newton and NuSTAR fit separating the XMM--Newton epochs.}
    \label{tab:app_jointFit}
    \begin{tabular}{lcccccc}
        \toprule
        \toprule
        \multirow{2}{*}{XMM epoch} & $N_H$ & \multirow{2}{*}{$\Gamma$} & $F_{3-10}$ & \multirow{2}{*}{$\chi^2/dof$} & \multirow{2}{*}{MOS constant} & \multirow{2}{*}{NuSTAR constant}\\
         & ($10^{23}\ \text{cm}^{-2}$) &  & ($10^{-12}\ \text{erg}\ \text{s}^{-1}\ \text{cm}^{-2}$) &  &  \\
        \midrule
        2000 & $2.29 \pm 0.15$ & $2.21 \pm 0.09$ & $3.4 \pm 0.2$ & 399.0/456 & $0.94 \pm 0.06$ & $1.01 \pm 0.06$ \\
        \midrule
        2003 & $2.21 \pm 0.11$ & $2.16 \pm 0.08$ & $3.6^{+0.1}_{-0.2}$ & 630.1/699 & $0.95 \pm 0.04$ & $0.92 \pm 0.04$ \\
        \bottomrule
    \end{tabular}
    \tablecomments{All uncertainties are reported at the 90\% confidence level.}
\end{table*}

\section{Additional SED plots} \label{app:sed_plots_extra}
\begin{figure*}[ht]
    \centering
    \begin{subfigure}{0.48\textwidth}
         \centering
         \includegraphics[width=\textwidth]{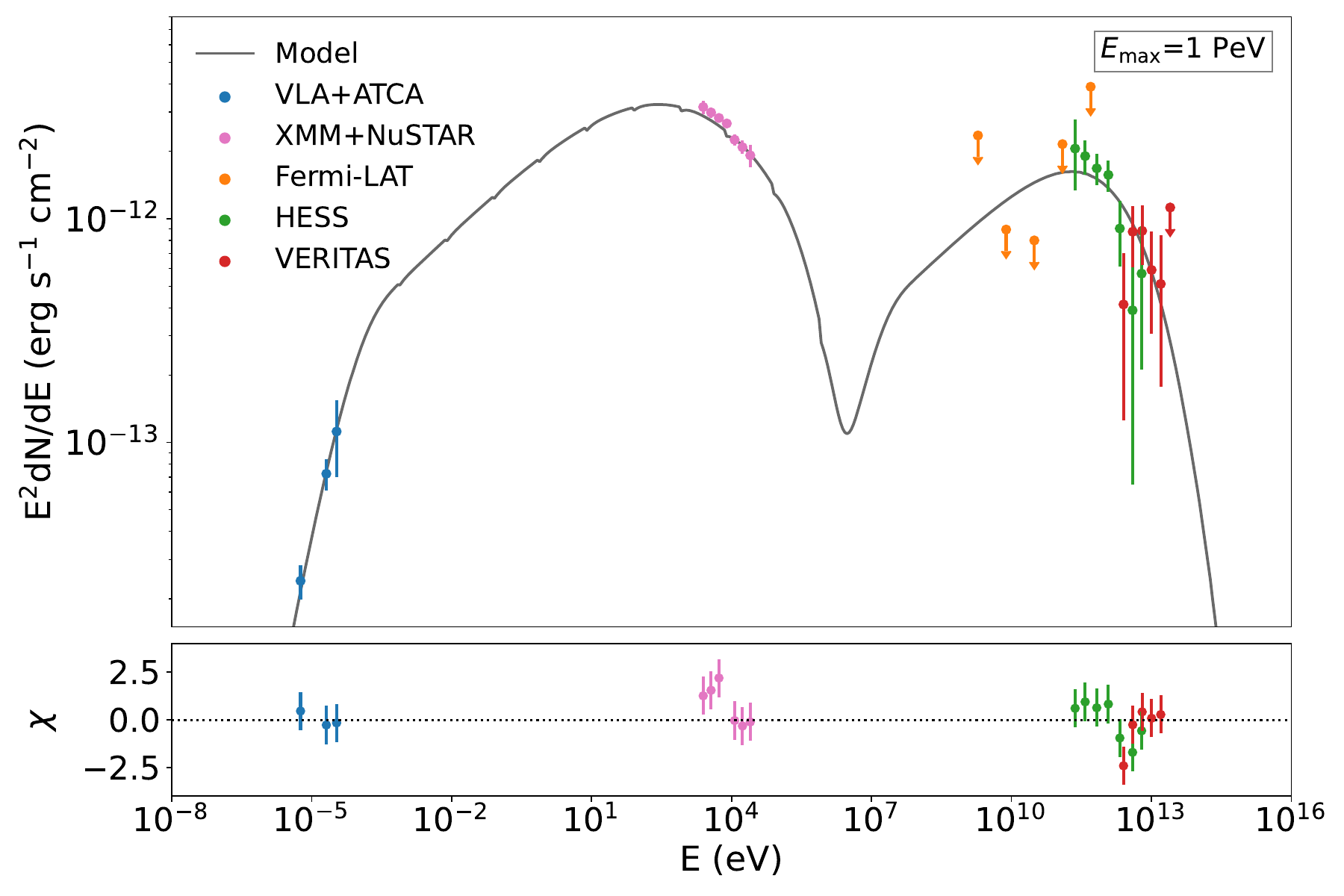}
         \caption{}
         \label{subfig:emax_1PeV_oneZ}
     \end{subfigure}
     \begin{subfigure}{0.48\textwidth}
         \centering
         \includegraphics[width=\textwidth]{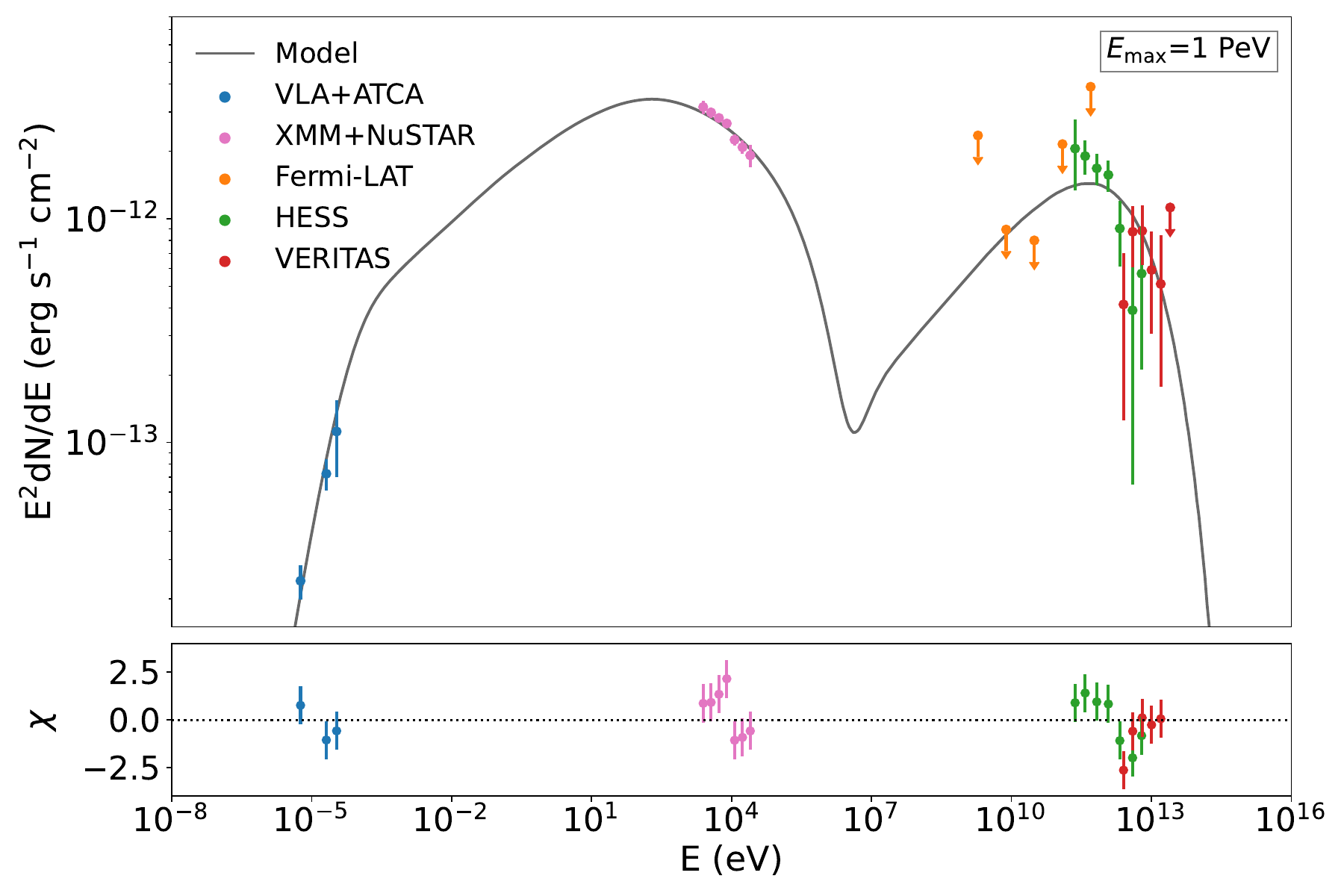}
         \caption{}
         \label{subfig:emax_1PeV_multiZ}
     \end{subfigure}
     \begin{subfigure}{0.48\textwidth}
         \centering
         \includegraphics[width=\textwidth]{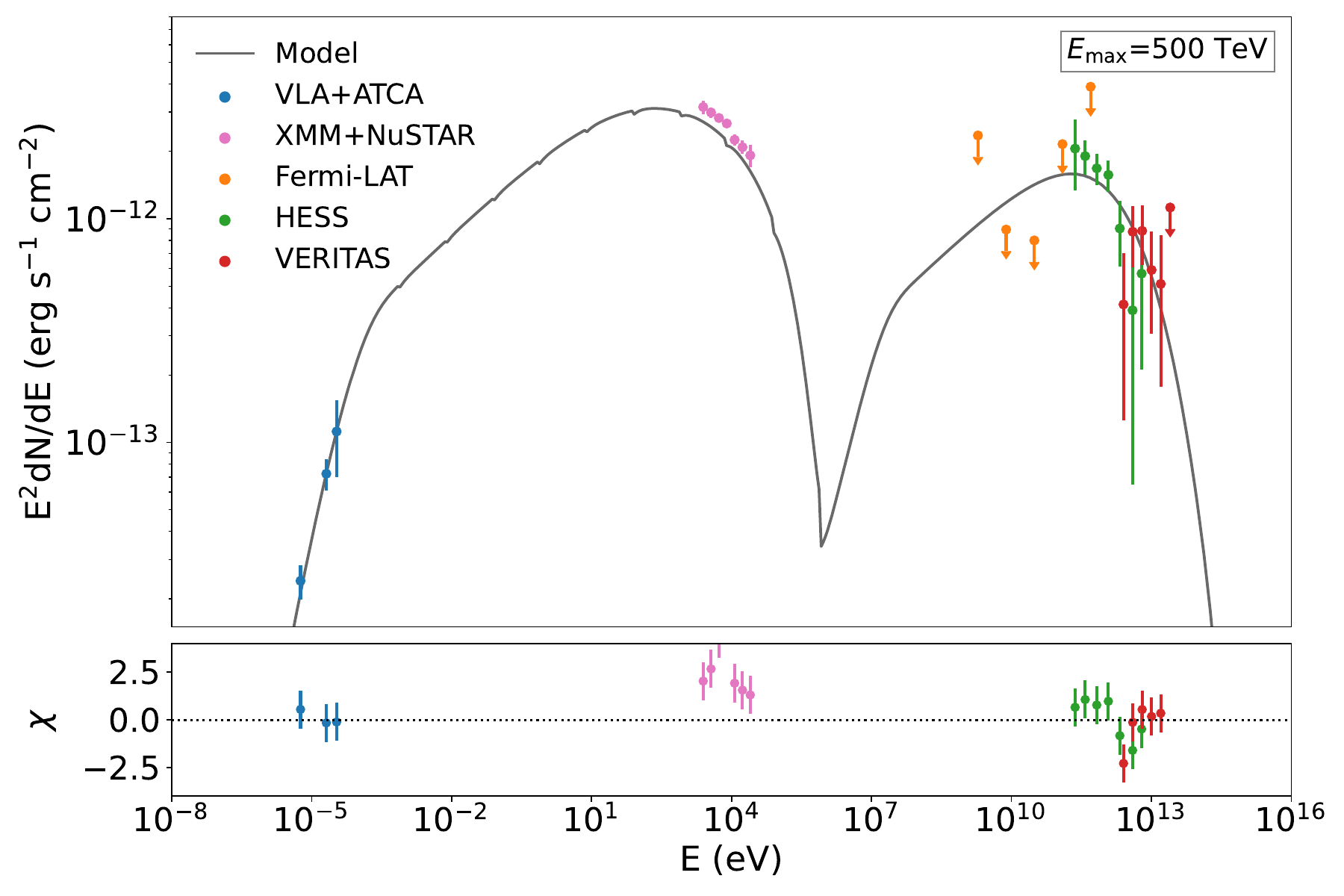}
         \caption{}
         \label{subfig:emax_500TeV_oneZ}
     \end{subfigure}     
     \begin{subfigure}{0.48\textwidth}
         \centering
         \includegraphics[width=\textwidth]{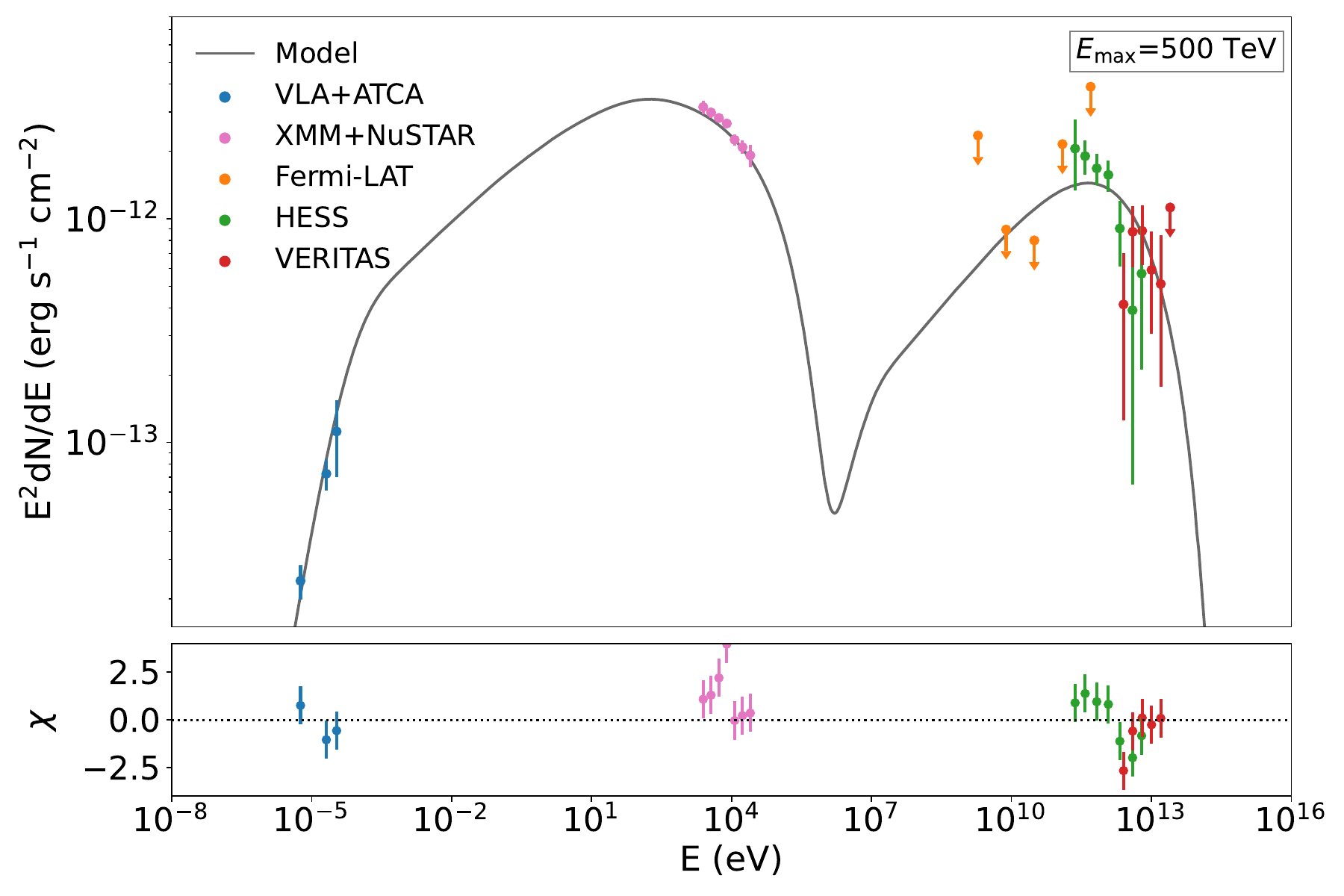}
         \caption{}
         \label{subfig:emax_500TeV_multiZ}
     \end{subfigure}
     \begin{subfigure}{0.48\textwidth}
         \centering
         \includegraphics[width=\textwidth]{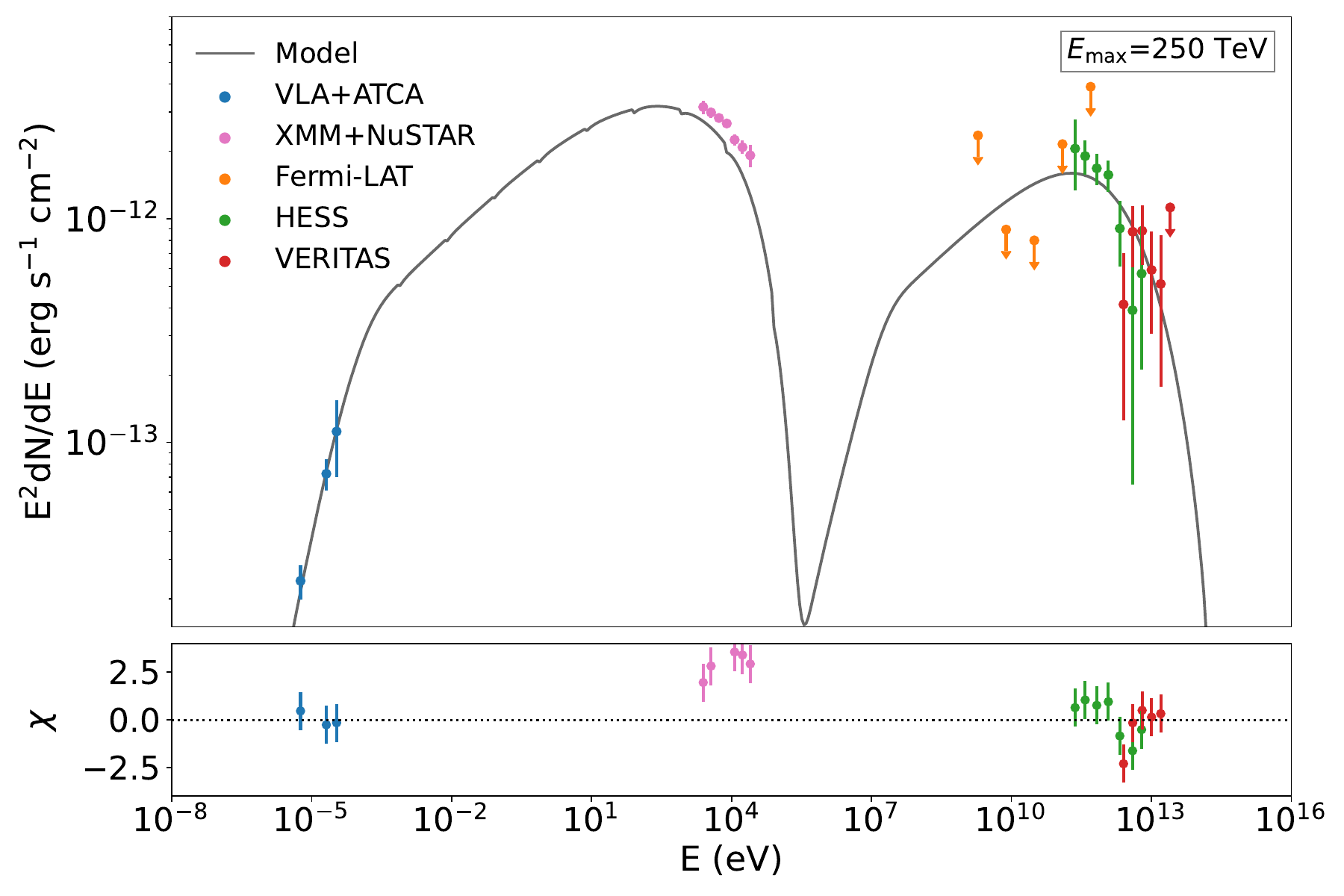}
         \caption{}
         \label{subfig:emax_250TeV_oneZ}
     \end{subfigure}
    \begin{subfigure}{0.48\textwidth}
         \centering
         \includegraphics[width=\textwidth]{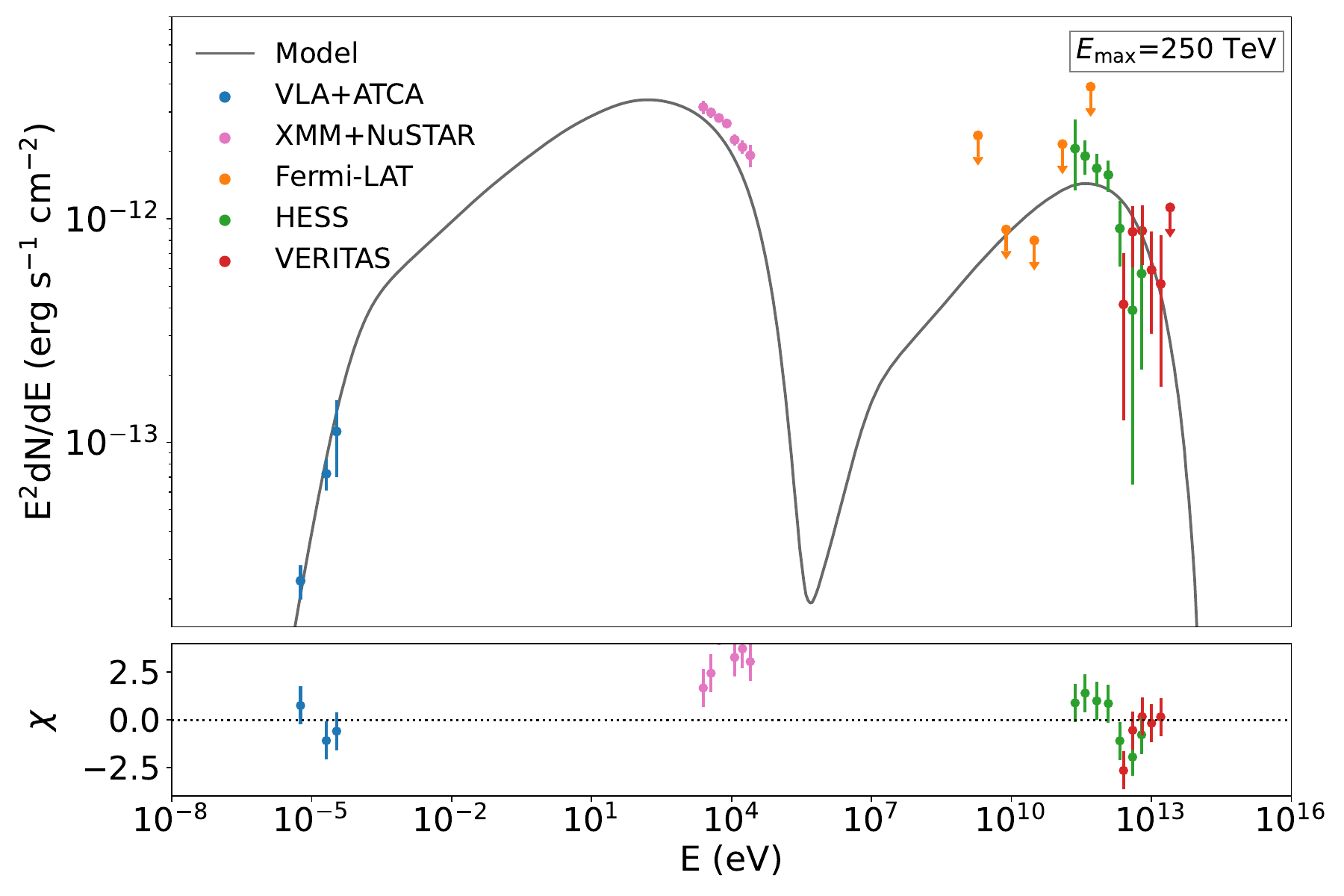}
         \caption{}
         \label{subfig:emax_250TeV_multiZ}
     \end{subfigure}
     \caption{Present-day SED of PWN~G0.9+0.1 fitted with the one-zone (left column) and multi-zone model (right column), decreasing the value of $E_{\rm max}$ to 1\,PeV (first row, panels (a) and (b) for one-zone and multi-zone), 500\,TeV (second row, panels (c) and (d) for one-zone and multi-zone), and 250\,TeV (third row, panels (e) and (f) for one-zone and multi-zone).}
    \label{fig:change_emax}
\end{figure*}

\begin{figure*}[ht]
    \centering
    \begin{subfigure}[b]{0.32\textwidth}
         \centering
         \includegraphics[width=\textwidth]{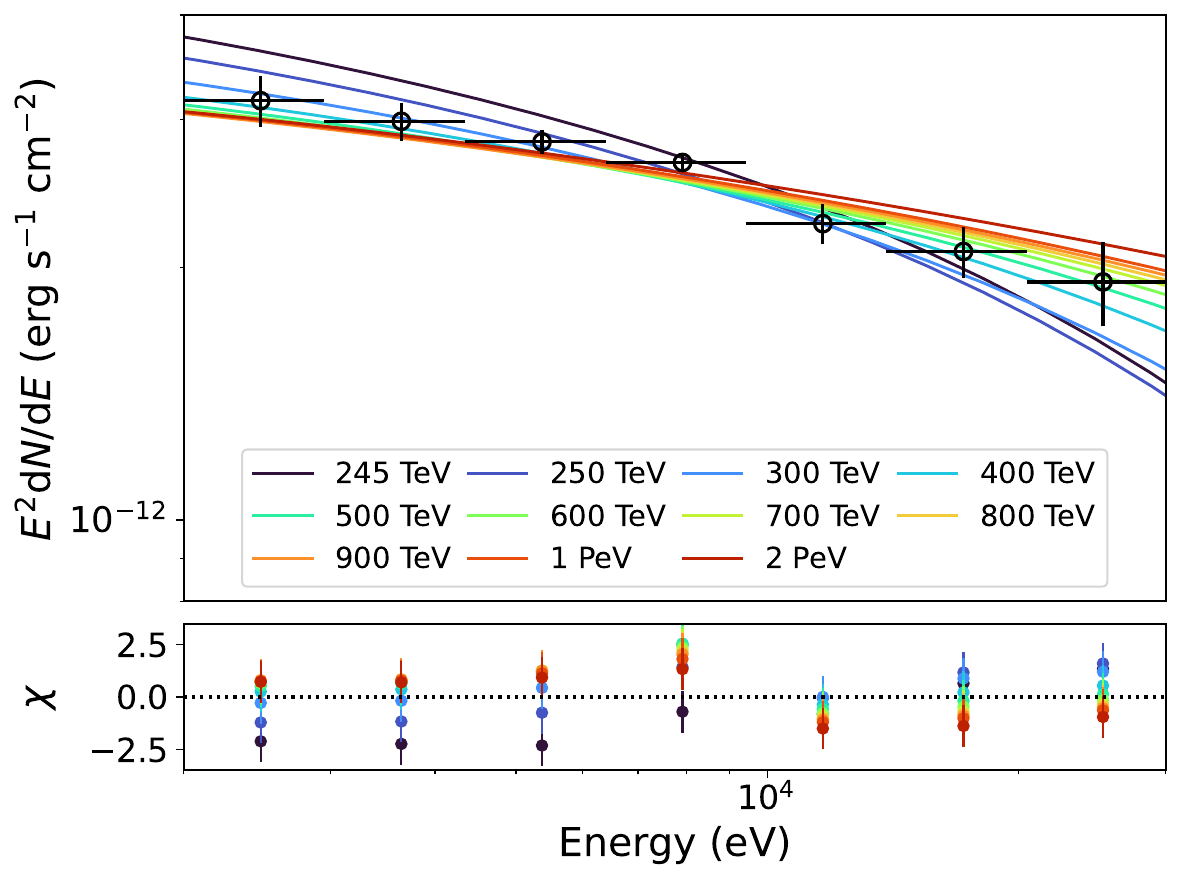}
         \caption{}
         \label{subfig:multiZone_Emax_sed}
     \end{subfigure}
     \begin{subfigure}[b]{0.32\textwidth}
         \centering
         \includegraphics[width=\textwidth]{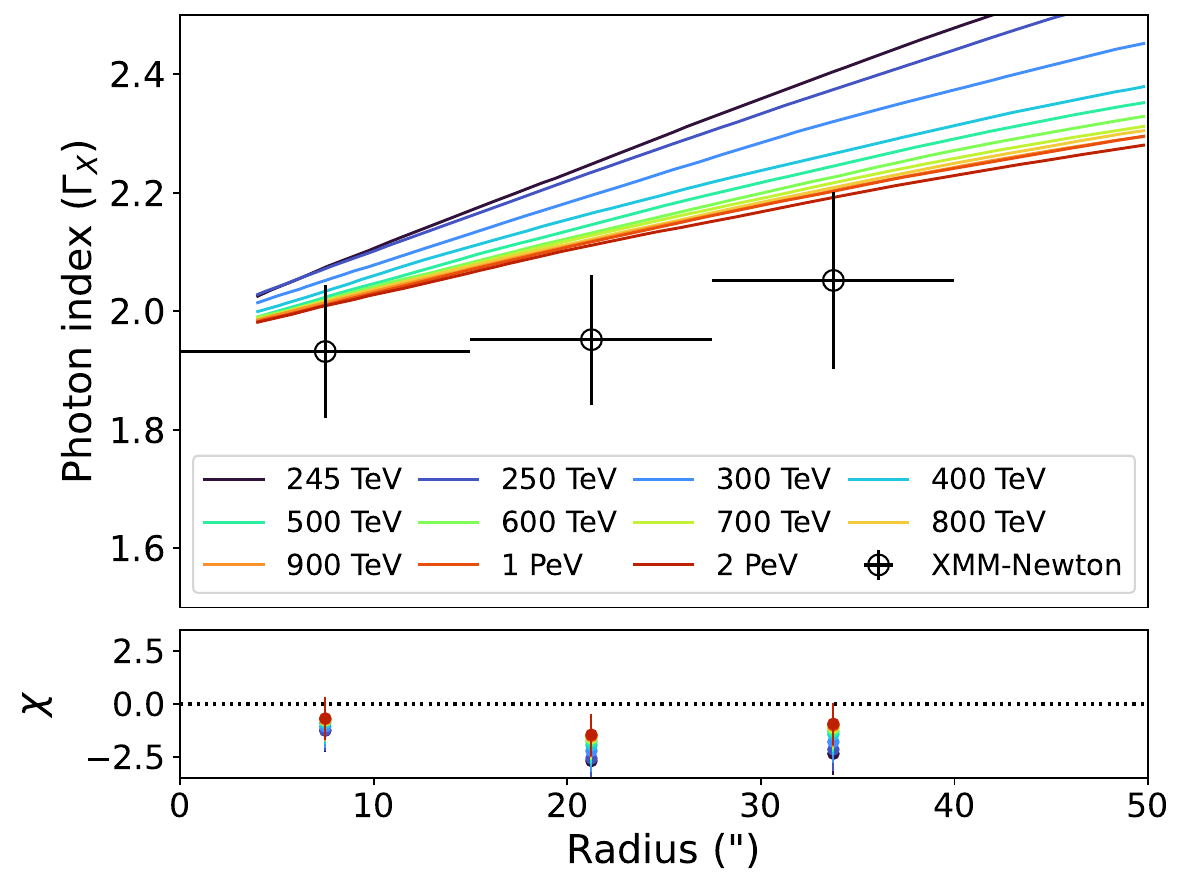}
         \caption{}
         \label{subfig:multiZone_Emax_gammaXmm}
     \end{subfigure}
     \begin{subfigure}[b]{0.32\textwidth}
         \centering
         \includegraphics[width=\textwidth]{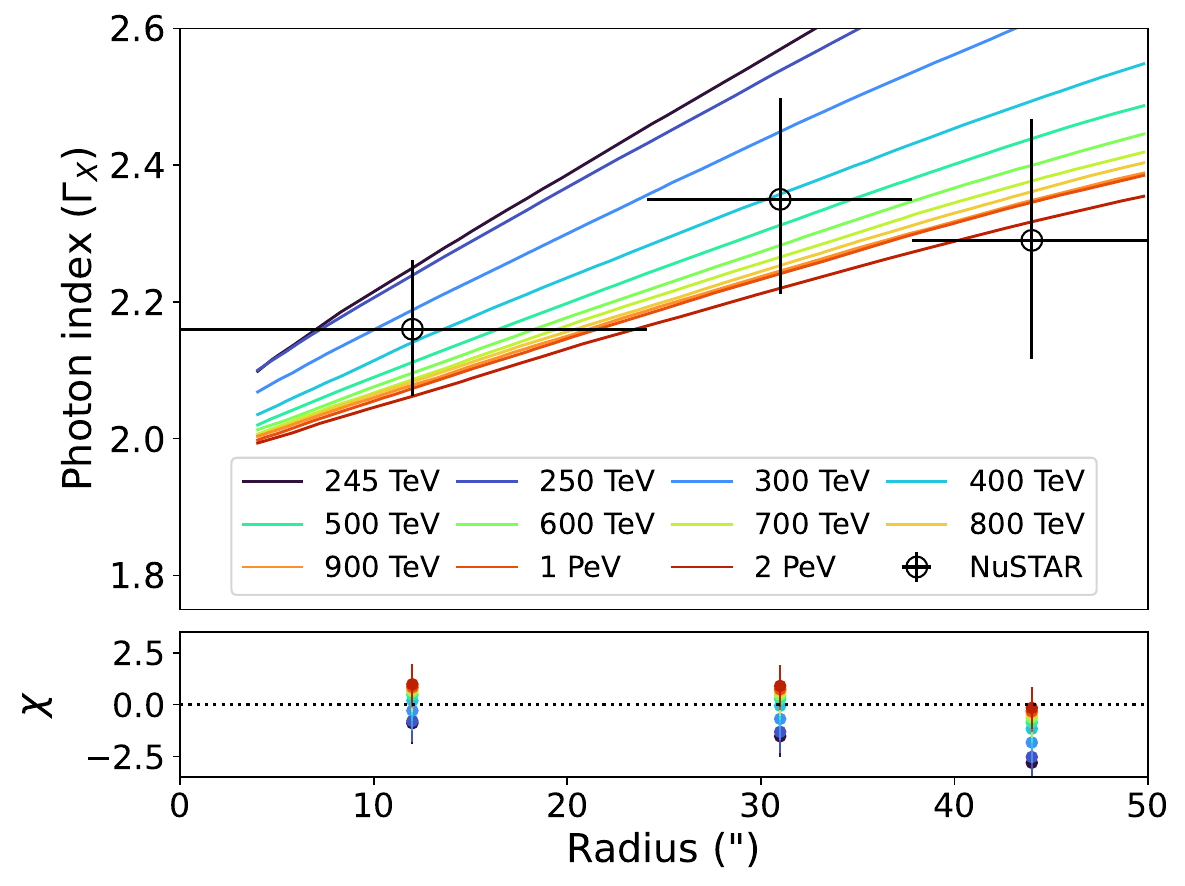}
         \caption{}
         \label{subfig:multiZone_Emax_gammaNu}
    \end{subfigure}    
    \caption{Results of the test for the robustness of the $E_{\rm max}$ estimate for the multi-zone model, for the present-day X-ray SED of PWN~G0.9+0.1 (a), for the XMM-Newton photon index profile (b), and for the NuSTAR photon index profile (c).}
    \label{fig:multiZone_test_Emax}
\end{figure*}

To further support our hypothesis that PWN~G0.9+0.1 could be a new PeVatron candidate, we produced the present-day SED plots for the one-zone and multi-zone models with progressively lower values of $E_{\rm max}$, leaving the other parameters fixed to their best-fit values of Tab.~\ref{tab:bestfit_bothModels}. The results of Fig.~\ref{fig:change_emax} show the progressive worsening of the residuals, especially in the X-ray band, as the maximum energy decreases. For the one-zone model, this approach is justified as $E_{\rm max}$ has been found to be not significantly correlated with any of the other parameters \citep{Gelfand_2015}. For the multi-zone model, instead, we also investigated the robustness of the $E_{\rm max}$ estimate by allowing other relevant parameters (i.e., $D_0$, $p_1$, $V_0$, $\alpha_V$, and $B_0$) to vary simultaneously in the fit. The results are shown in Figure \ref{fig:multiZone_test_Emax}. The results of this tests show that the fit quality degrades as $E_{\rm max}$ decreases from 2\,PeV ($\chi^2/dof =34/28$) to 245\,TeV ($\chi^2/dof =69/28$). In particular, for $E_{\rm max} < 400\,\mathrm{TeV}$, the value of $\chi^2$ rapidly increases with decreasing energy. This occurs because the cutoff energy in the model X-ray SED shifts to lower energies, leading to an under-prediction of the highest-energy NuSTAR data points. At the same time, the model-predicted photon index profile becomes softer, further worsening the agreement with the observed profile. We found that, in principle, the multi-zone model can accommodate $E_{\rm max}$ values down to 500\,TeV. However, $E_{\rm max}$ values above 1\,PeV are still statistically preferred for fitting both the X-ray SED and the photon index radial profiles.

\bibliography{bibliography}{}
\bibliographystyle{aasjournalv7}


\end{document}